%% file: main-arxiv.tex
\documentclass[sigconf,nonacm]{acmart}
\usepackage{lipsum} \usepackage{subcaption} \usepackage{macros}
\usepackage{graphicx} 
\usepackage{makecell}
\usepackage{subcaption}
\usepackage{csquotes}
\usepackage{fancybox}
\usepackage{wrapfig}
\usepackage{tcolorbox}
\usepackage{enumitem}
\usepackage{tabularx}

\newcommand{\hlcb}[2]{%
  \begingroup
  \setlength{\fboxsep}{0pt}
  \raisebox{0pt}[0.9ex][0.5ex]{\fcolorbox{#1!80!black}{#1!20}{\strut #2}}%
  \endgroup
}

\definecolor{ButtonColor}{HTML}{fcf8e8} 

\definecolor{listColor}{HTML}{944BF7}
\definecolor{clusterColor}{HTML}{54B9AF}
\definecolor{tableColor}{HTML}{D9885D}
\definecolor{timelineColor}{HTML}{54B9AF}
\definecolor{cartesianColor}{HTML}{FB6B6D}
\definecolor{graphColor}{HTML}{6289EE}

\definecolor{trajectory}{HTML}{FB6B6D}
\definecolor{projection}{HTML}{6289EE}

\definecolor{paperColor}{HTML}{3182C5}
\definecolor{hypothesisColor}{HTML}{8339EA}
\definecolor{experimentColor}{HTML}{E46124}
\definecolor{synthesisColor}{HTML}{883FE4}
\definecolor{findingColor}{HTML}{D47C22}

\AtBeginDocument{%
  \providecommand\BibTeX{{%
    \normalfont B\kern-0.5em{\scshape i\kern-0.25em b}\kern-0.8em\TeX}}}

\setcopyright{acmcopyright} \copyrightyear{2023} \acmYear{2023}
\acmDOI{10.1145/1122445.1122456}

\acmConference[Woodstock '23]{Woodstock '23: ACM Symposium on Neural
  Gaze Detection}{June 03--05, 2023}{Woodstock, NY}
\acmBooktitle{Woodstock '23: ACM Symposium on Neural Gaze Detection,
  June 03--05, 2023, Woodstock, NY}
\acmPrice{15.00} \acmISBN{978-1-4503-9999-9/18/06}

\begin{document}

\title[Atlas: Making Sense of Long Horizon Agent Trajectories]{Atlas: Making Sense of Long Horizon Agent Trajectories}


\author{Sangho Suh}
\affiliation{%
  \institution{Allen Institute for AI}
  \city{Seattle}
  \state{Washington}
  \country{United States}
}
\email{sanghos@allenai.org}

\author{Joseph Chee Chang}
\affiliation{%
  \institution{Allen Institute for AI}
  \city{Seattle}
  \state{Washington}
  \country{United States}
}
\email{josephc@allenai.org}

\author{Pao Siangliulue}
\affiliation{%
  \institution{Allen Institute for AI}
  \city{Seattle}
  \state{Washington}
  \country{United States}
}
\email{paos@allenai.org}

\author{Amy X. Zhang}
\affiliation{%
  \institution{University of Washington}
  \city{Seattle}
  \state{Washington}
  \country{United States}
}
\email{axz@cs.uw.edu}

\author{Matt Latzke}
\affiliation{%
  \institution{Allen Institute for AI}
  \city{Seattle}
  \state{Washington}
  \country{United States}
}
\email{mattl@allenai.org}

\author{Jonathan Bragg}
\email{jbragg@allenai.org}
\orcid{0000-0001-5460-9047}
\affiliation{%
  \institution{Allen Institute for AI}
  \city{Seattle}
  \state{WA}
  \country{USA}
}

\author{Doug Downey}
\affiliation{%
  \institution{Allen Institute for AI}
  \city{Seattle}
  \state{Washington}
  \country{United States}
}
\email{dougd@allenai.org}

\author{Daniel S. Weld}
\affiliation{%
  \institution{Allen Institute for AI}
  \city{Seattle}
  \state{Washington}
  \country{United States}
}
\email{danw@allenai.org}

\renewcommand{\shortauthors}{Suh et al.}

\newcommand{\sys}[0]{Atlas\xspace}

\begin{abstract}
AI agents can now search literature, generate hypotheses, analyze datasets, and run experiments at a scale difficult for researchers to match. Yet scaling scientific execution does not automatically scale researchers’ ability to make sense of what agents produce. When researchers return to a long-horizon run, they must evaluate work they did not monitor: what the agent produced, which decisions shaped those outputs, and whether those decisions served the research goal. We frame this post-run sensemaking as a representation problem across three linked levels: goal, decision, and artifact. From this framework we developed Atlas, an agentic workspace with a trajectory tree, which renders the run as a branching structure showing what the agent chose and set aside, and artifact projections over six representations. In a within-subjects study (N=14), participants reported higher decision transparency, artifact sensemaking, and goal connection with Atlas than with a standard baseline, together with substantially lower subjective workload.
\end{abstract}

\begin{teaserfigure}
  \centering
  \includegraphics[trim=0cm 0cm 0cm 0cm, clip=true, width=\textwidth]{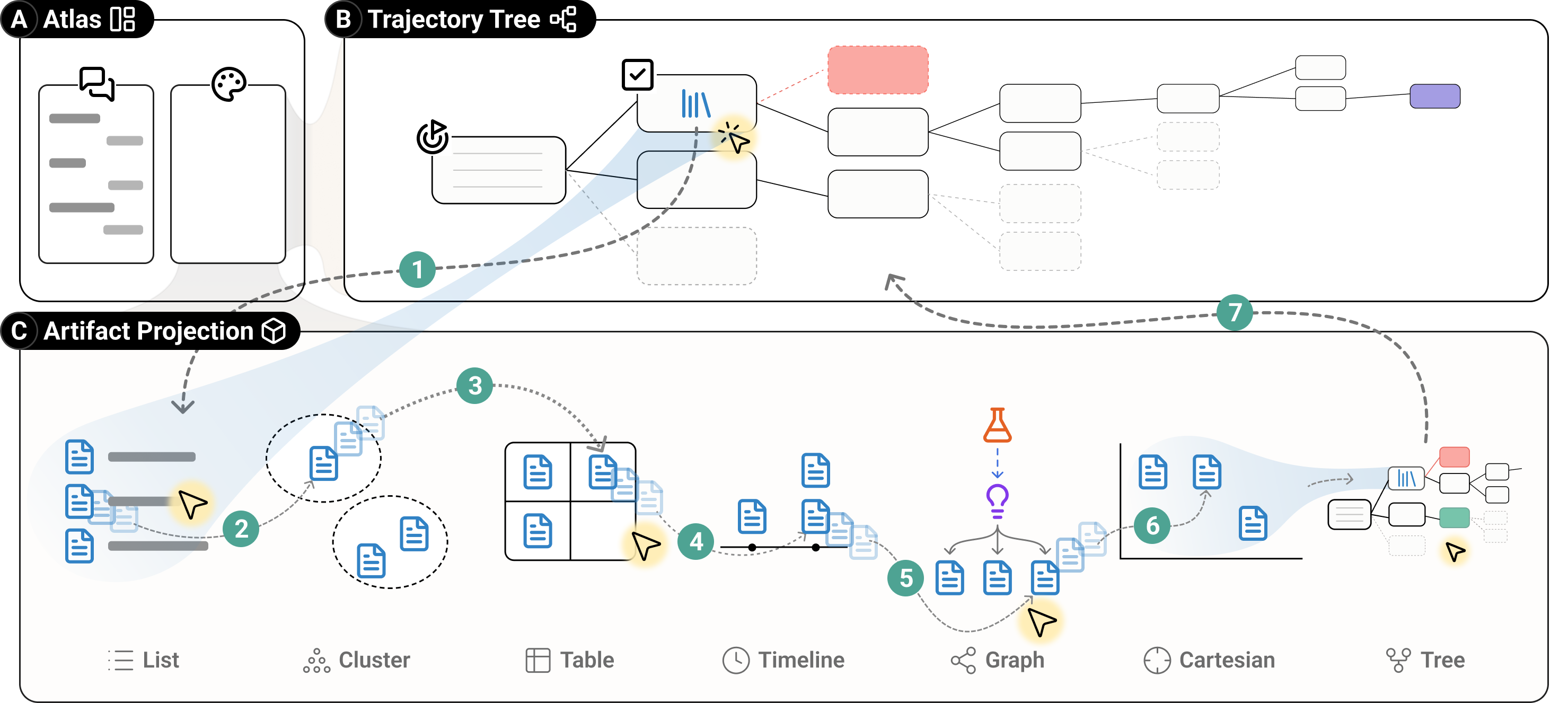}
  \caption{Atlas supports sensemaking of long horizon agent runs with (B) trajectory tree and (C) artifact projection. The trajectory tree exposes the \raisebox{-2pt}{\includegraphics[scale=0.4]{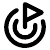}} research goal, \raisebox{-2pt}{\includegraphics[scale=1]{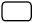}} pursued, \raisebox{-2pt}{\includegraphics[scale=1]{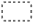}} deferred, and \raisebox{-2pt}{\includegraphics[scale=1]{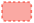}} ruled-out directions, the rationales behind these choices, and the artifacts (e.g., \raisebox{-2pt}{\includegraphics[scale=0.4]{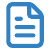}}, \raisebox{-2pt}{\includegraphics[scale=0.4]{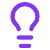}}, \raisebox{-2pt}{\includegraphics[scale=0.4]{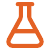}}) produced along the way. \raisebox{-2pt}{\includegraphics[scale=0.4]{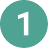}} Selecting an artifact set (e.g., \raisebox{-2pt}{\includegraphics[scale=0.4]{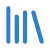}} paper set) in the trajectory opens its artifacts, which researchers can explore them using different representations (\raisebox{-2pt}{\includegraphics[scale=0.4]{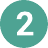}} $\rightarrow$ \raisebox{-2pt}{\includegraphics[scale=0.4]{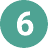}}) as their questions evolve: \textit{how do these papers compare?} $\mapsto$ \textsf{table}; \textit{when were they published?} $\mapsto$ \textsf{timeline}; \textit{how are they related to hypotheses and experiments?} $\mapsto$ \textsf{graph}. Researchers can also traverse between artifacts and the decisions that produced them (e.g., $\Downarrow$ \raisebox{-2pt}{\includegraphics[scale=0.4]{figures/icons/1-icon.pdf}} $\cdot$ $\Uparrow$ \raisebox{-2pt}{\includegraphics[scale=0.4]{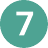}}).}
  \Description{Teaser figure}
  \label{fig:teaser}
\end{teaserfigure}

\ccsdesc[500]{Human-centered computing~Interactive systems and tools}
\ccsdesc[300]{Human-centered computing~Empirical studies in HCI}

\keywords{human-agent interaction, long-horizon tasks, agent trajectory, sensemaking, provenance, representations}

\maketitle

\section{Introduction}
\label{sec:intro}

\begin{displayquote}
    ``\textit{The purpose of computing is insight, not numbers}.'' — Richard Hamming
\end{displayquote}

More than six decades ago, Hamming opened his book \textit{Numerical Methods for Scientists and Engineers} with this reminder that the value of computation lies not simply in what it \textit{produces}, but in what it helps us \textit{understand}. This reminder is even more important today as AI agents take on an increasingly large portion of scientific work and \textit{produce} scientific outputs at remarkable speed and scale. Agents can now find papers, generate hypotheses, execute experiments, and produce findings~\cite{jansen2025codescientist, pu2025ideasynth, agarwal2026autodiscovery}. Yet our focus on accelerating scientific execution risks overlooking an equally important challenge: helping researchers make sense of the resulting work, including how findings were produced, how different pieces relate to one another, and what they mean for researchers’ own goals. Without comparable support for sensemaking, the pace of agents’ scientific output can outstrip researchers’ capacity to process it, leaving valuable insights unrealized.


This problem becomes more pronounced in long-horizon agent workflows, where researchers delegate complex work that requires many steps and intermediate decisions and return hours or days later to a completed run they did not continuously observe and steer. 
In this setting, researchers need to recover the parts of the run that matter for interpreting its outcomes: consequential decisions, relationships among outputs, and connections to the research goal.
Unlike in-run interaction, where researchers can inspect and redirect the agent as decisions are being made, post-run sensemaking begins after those decisions have already accumulated across hours of execution and many heterogeneous outputs.

Consider an AI researcher who asks an AI agent to investigate \textit{why chain-of-thought prompting improves accuracy on some reasoning benchmarks but not others}. They return the next morning to a run in which the agent has processed more than \raisebox{-2pt}{\includegraphics[scale=0.4]{figures/icons/artifact-paper.pdf}} 120 papers, generated \raisebox{-2pt}{\includegraphics[scale=0.4]{figures/icons/artifact-hypothesis.pdf}} 12 hypotheses, run \raisebox{-2pt}{\includegraphics[scale=0.4]{figures/icons/artifact-experiment.pdf}} 6 experiments across several models, and produced \raisebox{-2pt}{\includegraphics[scale=0.4]{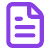}} 6 reports. Along the way, the agent narrowed six model families to two and adopted a scoring threshold that discarded partially correct answers. 
The researcher may not need to read every intermediate step, but understanding these choices is essential to interpreting the findings and advancing their understanding of the topic. However, recovering these decisions from a transcript requires locating relevant moments across hours of output and mentally connecting the decisions to one another and to the findings.

Asking the agent for a summary or follow-up explanation only partly addresses this problem. Such interaction is useful when researchers know what they want to retrieve, but post-run sensemaking also involves discovering what warrants attention in the first place: which decisions were consequential, which directions were considered but abandoned, which artifacts relate or conflict, and where the run may have departed from the researcher's intent. A sequence of follow-up questions therefore still leaves the researcher responsible for reconstructing these relationships across the run.

Recent work has begun to make agent executions more inspectable by reconstructing chronological traces into provenance graphs, structured reasoning records, or specialized audit views. These approaches demonstrate the value of moving beyond raw execution logs, but primarily target provenance capture, debugging, cross-run analysis, or post-hoc auditing~\cite{wu2026agenttrails, vispute2026reasoning, macko2011provenance}. We ask a complementary question: \textbf{what representations and interactions do researchers need to make sense of long horizon agent runs?} Rather than confining provenance to a dedicated trace or provenance view, we investigate how provenance relationships can serve as an underlying structure connecting multiple representations of a run as researchers move among questions about the agent's goals, decisions, and artifacts, as illustrated in Fig~\ref{fig:teaser}.

To understand how researchers currently perform this work, we conducted a formative study with 12 researchers and research engineers who use agents for research tasks lasting from minutes to many hours and even days (Sec~\ref{sec:formative-study}). 
All participants described difficulty recovering what an agent had decided and why, as well as making sense of the outputs it produced across a run. Participants also developed their own ways to track connections across outputs, decisions, and research goals.
Overall, these difficulties clustered into three challenges: (1) \textbf{decision transparency}, understanding what the agent chose and why; (2) \textbf{artifact sensemaking}, synthesizing across what the run produced; and (3) \textbf{goal provenance}, connecting those decisions and outputs back to what the researcher was trying to investigate.

Together with Russell et al.'s account of sensemaking as a search for representations~\cite{russell1993cost}, we used these findings to derive a framework for post-run sensemaking across three linked levels of an agent run: goal, decision, and artifact (Sec.~\ref{sec:framework}). 
We argue that researchers need stable references to research goals, visibility into the structure of the agent’s decisions, and multiple representations for interpreting artifacts. Provenance connects these levels, allowing researchers to trace how decisions and artifacts relate to their goals.

Atlas instantiates this framework as an agentic research workspace in which artifacts and the agent’s decision trajectory coexist as persistent, linked objects (Sec~\ref{sec:system}). Instead of exposing provenance through a single dedicated view, Atlas lets researchers project the same artifact set into multiple coordinated representations and traverse from those artifacts to the decisions that produced them and back to the research goal. 
This supports two complementary interactions during post-run sensemaking: switching representations as questions change and following provenance links among goals, decisions, and artifacts.

We evaluated Atlas in a within-subjects study with 14 researchers and an exploratory deployment with 6 participants. In the controlled study, participants reviewed two long-running research runs in counterbalanced interface conditions, one using Atlas and one using a standard chat-and-artifact workspace. Each interface provided access to the complete transcript and artifacts associated with its run; Atlas additionally provided the trajectory tree and multi-representational canvas. Participants reported higher decision transparency, artifact sensemaking, and goal connection with Atlas, together with substantially lower subjective workload. We additionally use the deployment to examine how these mechanisms were used in participants’ own research and where the workspace still fell short.

\begin{itemize}
    \item \textbf{A formative study (N=12)} that outlines three challenges of post-run sensemaking over long-horizon agent runs: decision transparency, artifact sensemaking, and goal provenance.
    \item \textbf{A representation framework for post-run sensemaking} that characterizes agent runs across goal, decision, and artifact levels, derives representational requirements at each level, describes how provenance links enable traversal among them.
    \item \textbf{\sys}, an agentic research workspace that instantiates this framework through a trajectory view and multiple coordinated artifact representations, linked by provenance.
    \item \textbf{A within-subjects evaluation (N=14) and exploratory deployment study (N=6)} examining how Atlas supports post-run sensemaking. In the controlled study, participants, participants reporting higher decision transparency, artifact sensemaking, and goal connection, together with lower subjective workload than with a standard chat-and-artifact workspace. The deployment further examined how participants used Atlas across sessions and identified limitations that emerged through continued use.
\end{itemize}

\section{Related Work}
\label{sec:related-work}

Our work sits at the intersection of autonomous systems for scientific research, human-agent interaction around long-horizon tasks, provenance and trajectory visualization, and tools for research sensemaking. We review each area in turn and distinguish Atlas from systems primarily concerned with improving agent execution, supporting in-run oversight, or capturing and inspecting structured execution traces.

\subsection{Agentic Systems for Scientific Discovery}

Recent years have seen a surge of AI systems capable of operating autonomously across extended research pipelines. The AI Scientist~\cite{lu2024aiscientist} and Agent Laboratory~\cite{schmidgall2025agentlab} automate the research cycle end to end, from literature search and hypothesis generation through experiment execution and manuscript writing, without continuous human involvement. Among domain-specific agents, Co-Scientist~\cite{gottweis2026accelerating} applies a multi-agent generate-debate-evolve loop to biomedical hypothesis generation, Biomni~\cite{huang2025biomni} equips agents with curated biomedical tool libraries for generalist scientific tasks, and CodeScientist~\cite{jansen2025codescientist} chains code-based experimentation into semi-automated discovery workflows. Together, these systems demonstrate that agents can increasingly carry out multi-stage scientific work over extended periods.

Much of this work has focused on expanding what agents can autonomously accomplish. As agents take on longer and more consequential portions of the research process, a complementary question becomes increasingly important: how can researchers understand, evaluate, and act on work that an agent performed in their absence? This requires more than assessing a final output, because a completed run may contain consequential decisions, abandoned directions, intermediate evidence, and heterogeneous artifacts whose relationships determine how the result should be interpreted. Atlas focuses on this human side of delegated scientific work: how researchers make sense of a completed exploratory run, evaluate whether its decisions served their research goal, and determine what to pursue next.

\subsection{Human-Agent Interaction Before and During Long-Horizon Tasks}

A growing body of work examines how humans can maintain agency over long-running tasks before or while agents execute them. Cocoa~\cite{feng2026cocoa} introduces co-planning and co-execution as interaction primitives for scientific research, allowing humans and agents to jointly construct plans and interleave execution. Magentic-UI~\cite{mozannar2025magentic} operationalizes mechanisms such as co-tasking, action guards, and answer verification for general-purpose web agents. Both support interactive oversight in which the human remains present and can steer the agent as work unfolds.

Other work examines how people specify or monitor goals. Compass versus Railway Tracks~\cite{petridis2026compass} shows that people tend to communicate high-level intent when delegating to humans but provide AI systems with more exhaustive, script-like instructions because they are less willing to trust the system to infer intent and prioritize. Kazemitabaar et al.~\cite{kazemitabaar2024decomposition} use interactive task decomposition to separate assumptions from actions and make generated data-analysis workflows editable and verifiable. OnGoal~\cite{coscia2025ongoal} provides real-time feedback about goal alignment in multi-turn LLM interactions, while OrchVis~\cite{zhou2025orchvis} represents goals explicitly and monitors their progress in multi-agent workflows.

These systems primarily support specifying intent before execution or steering agents while work unfolds. Atlas focuses on the researcher’s return after an extended period of autonomous execution, when they must determine which decisions matter, how those decisions shaped the outputs, and whether the run advanced their research goal. Post-run sensemaking can also inform subsequent oversight by helping researchers identify where to resume or redirect the agent.


\subsection{Agent Provenance, Trajectory, and Execution Visualization}

Provenance and visualization research has long studied how to make computational processes inspectable. Scientific workflow systems such as VisTrails capture how data and computational steps produce an outcome, while visualization systems such as Provenance Map Orbiter, InProv, and AVOCADO explore how provenance graphs can be summarized, reorganized, or navigated at multiple levels of detail~\cite{bavoil2005vistrails, macko2011provenance, borkin2013evaluation, stitz2016avocado}. More broadly, work on coordinated multiple views has shown that different representations can expose complementary properties of the same underlying data~\cite{roberts2007state}. 
Atlas builds on these approaches by linking multiple artifact representations to the agent’s decision trajectory and research goals to support post-run sensemaking.

Recent work extends these ideas to AI agents. PROV-AGENT models agent actions, prompts, model invocations, data, and downstream workflow tasks within a unified provenance structure, while reasoning-provenance approaches capture structured records of intent, observation, inference, and plan revision for downstream analysis~\cite{souza2025provagent, vispute2026reasoning}. Other work uses trajectories primarily for diagnosis. AgenTracer, for example, identifies the agent and the decisive step responsible for failures in multi-agent executions~\cite{zhang2026agentracer}. Related user-facing work on computer-use agents presents coordinated views of actions, resources, permissions, and provenance to support post-hoc risk auditing and remediation~\cite{peng2026what}.

The closest recent systems to Atlas are AgentTrails and Graph of Trace. AgentTrails transforms chronological logs into provenance graphs that expose dependencies among tool calls and artifacts, and supports overview, single-trace inspection, and comparison across executions for tasks such as debugging, pattern extraction, and workflow reuse~\cite{wu2026agenttrails}. Graph of Trace similarly represents scientific agents, constructing a directed graph of intermediate research steps and dependencies so users can monitor and inspect an execution as it unfolds. Its expert evaluation suggests that structured traces improve interpretability and usability relative to reviewing raw dialogue and logs~\cite{gao2026graph}.

Atlas shares the premise that chronological logs are a poor representation of complex agent execution, but takes a different starting point. Rather than asking primarily how to capture, diagnose, or visualize an execution trace, we ask what representations researchers need to make sense of a completed exploratory run. Our contribution is not simply to expose provenance, whether through links, graphs, or structured traces, but to use provenance relationships as an underlying structure connecting multiple representations. Researchers must reason about the run’s goal, the agent’s decisions, and the heterogeneous artifacts it produced, and the appropriate representation changes with the question being asked. Atlas therefore allows researchers to shift among representations of the same artifact set while tracing artifacts to the decisions that produced them and back to the research goal.

\subsection{Research Sensemaking and Multiple Representations}

Our approach to post-run sensemaking is grounded in work on sensemaking and external representations. Russell et al.~\cite{russell1993cost} characterize sensemaking as a search for a representation that makes task-specific operations easier: people generate a representation, instantiate information within it, encounter residue that does not fit, and shift to another representation as needed. The costs of generating and shifting representations therefore shape which forms of analysis people attempt at all.

Research tools have long externalized parts of scholarly sensemaking to lower sensemaking costs. One line of work augments individual papers with citation context, progressive disclosure, related media, and other information that reduces the cost of entering and understanding a document~\cite{fok2024qlarify,kim2023papeos,lo2023semantic, chang2023citesee}. Another helps researchers construct relationships across papers. Threddy and Synergi organize literature into threads and syntheses~\cite{kang2022threddy, kang2023synergi}, while Relatedly~\cite{palani2023relatedly}, PaperWeaver~\cite{lee2024paperweaver}, DiscipLink~\cite{zheng2024disciplink}, and related systems contextualize newly encountered papers against an existing collection~\cite{newman2024arxivdigestables, zhang2008citesense, fok2025facets, duck2025finding, martin2026papertrail, chang2026asterism}. 

Other systems externalize these structures spatially. Garden of Papers~\cite{ma2025garden}, Argo Scholar~\cite{li2021argo}, and ScholarMate~\cite{ye2025scholarmate} provide spatial organizations of paper collections; Graphologue turns LLM responses into interactive diagrams~\cite{jiang2023graphologue}; Sensecape lets users organize generated information hierarchically across levels of abstraction~\cite{suh2023sensecape}; and CrossLit coordinates a canvas of research materials with a developing manuscript~\cite{choe2026crosslit}, and Jigsaw is particularly relevant in coordinating multiple representations of entities and relationships extracted from a text collection~\cite{stasko2008jigsaw}.


These systems demonstrate the value of external representations for scholarly sensemaking, but most assume a collection that the researcher gathered or incrementally constructed and whose provenance they therefore largely understand~\cite{choe2024fields, zhu2025sense}. They also generally operate over a relatively \textit{homogeneous} artifact type, most often papers (e.g., \cite{ma2025garden}) or notes (e.g., \cite{duck2025finding}). A long-horizon research agent changes both conditions. 
An agent can retrieve papers, generate hypotheses, execute experiments, and produce analyses and syntheses, while making decisions that they did not directly observe but shape which artifacts are created and how they relate.
In our formative study, participants constructed matrices, dependency graphs, task graphs, and other structures to address different questions about agent runs.




Atlas builds on prior work by supporting representational flexibility over a persistent, \textit{heterogeneous} artifact set. The same artifacts can be projected into different coordinated representations as the researcher’s question changes (e.g., cluster $\mapsto$ table) , while a separate structural representation exposes consequential agent decisions. Provenance connects these representations, allowing researchers to trace a finding to the decision that produced it and back to the research goal, or to follow relationships among papers, hypotheses, experiments, and findings. In this way, Atlas supports two complementary forms of movement during post-run sensemaking: representational shifts within the artifact level and provenance-based traversal across levels of the run.

\section{Formative Study}
\label{sec:formative-study}

\subsection{Method}

We conducted a formative interview study with 12 researchers and research engineers working in AI/ML, NLP, HCI, economics, and neuroscience. All used AI coding agents on a daily or near-daily basis, running agents on complex research tasks with or without fully pre-specified plans. Table~\ref{tab:participants} summarizes participant backgrounds. Participants reported agent runs lasting from under 30 minutes to over 19 hours. We conducted semi-structured interviews over video call (40--70 minutes each), supplemented by a pre-study background survey. Transcripts were analyzed through iterative open coding. Participants are referred to as F1--F12. The study received IRB approval, and participants were compensated with a \$60 gift card.

\begin{table*}
  \centering
  \footnotesize
\caption{Formative study participant backgrounds.}
  \label{tab:participants}
  \resizebox{\linewidth}{!}{
  \begin{tabular}{@{}lllllll@{}}
    \toprule
    \textbf{ID} & \textbf{Role} & \textbf{Research Domain} & \textbf{Experience} 
      &  \makecell[l]{\textbf{Typical Run Duration}} & \textbf{Output Types} &
      \textbf{Sensemaking Time} \\
    \midrule
    F1  & Researcher (PhD) & AI/ML (NLP) & 6--10 yrs          & 30 min--4 hr & Reports, code, plots        & 1--2 hr       \\
    F2  & Research engineer     & AI/ML (Systems) & 6--10 yrs            & 1--8 hr      & Code, test results          & 30 min--2 hr  \\
    F3  & Researcher    & AI/ML (HCI) & 3--5 yrs                  & 1--4 hr      & Code, transcripts, reports  & 1--3 hr       \\
    F4  & Researcher (PhD)       & Computational Neuroscience  & 6--10 yrs  & 4--19+ hr    & Reports, figures            & 2+ hr         \\
    F5  & Researcher (PhD)      & AI for Science   & 6--10 yrs            & 1--8 hr      & Code, analysis outputs      & 1--2 hr       \\
    F6  & Researcher   & ML / Neuroscience  & 3--5 yrs           & 5--30 min    & Diagrams, code, plots       & 30 min--1 hr  \\
    F7  & Researcher (PhD)        & Systems  Neuroscience  & 6--10 yrs      & 2--8 hr      & QC plots, figures           & 2+ hr         \\
    F8  & Researcher         & HCI / Behavioral Economics  & 3--5 yrs& 1--4 hr      & Text, analysis outputs      & 1--2 hr       \\
    F9  & Research engineer & HCI / AI Systems  & 3--5 yrs            & 2--8 hr      & Task graphs, PDF reports    & 1--2 hr       \\
    F10 & Researcher (PhD)   & Computational Neuroscience    & 3--5 yrs  & 5--19+ hr    & Reports, model weights      & 2+ hr         \\
    F11 & Researcher         & NLP / HCI    & 6--10 yrs                 & 1--4 hr      & Figures, reports, code      & 1--2 hr       \\
    F12 & Researcher         & Retina Neuroscience     & 10+ yrs       & 1--4 hr      & Figures, HTML artifacts     & 30 min--1 hr  \\
    \bottomrule
  \end{tabular}
  }
\end{table*}

\subsection{Findings}
\label{sec:formative-findings}

\subsubsection{\textbf{Challenge 1: Goal Provenance --- Difficulty connecting agent outputs to the research goal.}}

Participants described difficulty determining how agent outputs advanced their research goals. They developed their own ways to preserve these connections and assess whether the agent’s work was relevant and worth pursuing.

F9 developed a task-graph system that allowed each task to be checked against its originating specification, not just traced back to its code. F4 asked the coding agent to visualize the session history with clickable decision points for revisiting earlier moments in the run. F10 used a markdown file (\code{mission.md}) to encode research objectives and analytical preferences, which the agent was instructed to consult during every review. F12 used a knowledge-graph tool to preserve relationships among supporting and contradicting hypotheses across sessions.

Participants reported that written plans did not fully specify the choices agents made during execution. Plans constrained the overall task but left open which directions to pursue, which hypotheses to prioritize, and how to resolve ambiguities. Researchers therefore had to reconstruct how these choices related to their research goals after the run.

\begin{tcolorbox}[title=Challenge 1: Goal Provenance] 
Researchers struggle to trace how the agent’s decisions and outputs connect to their research goals. Understanding what the agent did is insufficient for evaluating whether its choices advanced those goals; researchers also need explicit connections among artifacts, decisions, and research intent.

\textbf{Design implication:} 
Keep research goals accessible during sensemaking and link them to decisions and artifacts, so researchers can trace how the agent’s work addressed what they were trying to investigate.
\end{tcolorbox}

\subsubsection{\textbf{Challenge 2: Decision Transparency --- Difficulty reconstructing what the agent decided and why.}}

All 12 participants described difficulty reconstructing what the agent chose and why. 
These difficulties concerned both strategic decisions about research direction and implementation choices that shaped the results.

At the strategic level, agents made consequential choices about research direction without surfacing them. F1 observed the agent discarding promising hypotheses with no visible reasoning. F3 described silent choices as ``\textit{forks in the road where [agent] could have made five different decisions here and [it] picked one and that was a matter of taste and [agent] doesn't have taste, right? so [it] ha[s] to tell me about that.}'' F2's experience illustrates why verifying adherence to a high-level plan was insufficient: even after confirming that the agent followed the intended plan, they later discovered an implementation-level error that affected the result. 
At the implementation level, F2 caught errors only by noticing implausible assumptions; F8 accidentally discovered a hardcoded sampling limit of N=20 instead of N=2000 in the experiment design; F5 described a single incorrect \code{if-statement} invalidating an entire experiment, caught only because the participant was a domain expert that could tell the result was suspicious.

The invisibility of decisions had a direct effect on delegation. 
Every participant described delegating more when they could ``see'' that agent decisions were relevant and productive for their research question, and pulling back when they could not. 
When the agent sometimes surfaced a set of potential decisions in a structured manner with explicit rationale to help F4 reason about the decision, F4 reported following the agent's top recommendation 80\% of the time. The converse was equally claer. F3 described how the lack of visibility into the agent's decisions and getting ``\textit{bit hard}'' as a consequence reduced his willingness to delegate long-running tasks to the agent in future sessions.

\begin{tcolorbox}[title=Challenge 2: Decision Transparency] Researchers struggle to reconstruct the agent’s strategic decisions, including directions pursued or abandoned, and implementation choices, including constraints and assumptions. This limits their ability to evaluate the work and can reduce their willingness to delegate.


\textbf{Design implication:} Make decisions and their rationale inspectable at both levels, so researchers can assess how individual choices shaped the results and whether they were appropriate for the research goal.

\end{tcolorbox}

\subsubsection{\textbf{Challenge 3: Artifact Sensemaking --- Difficulty synthesizing across agent outputs.}}

All participants described difficulty making sense of agent outputs and reported often or always feeling overwhelmed by their volume. Their accounts also revealed difficulty synthesizing across artifacts: identifying which findings related to which hypotheses, which results were consistent or contradictory, and what patterns across the run suggested for future research directions.


Participants described building their own visualizations and organizational tools to make sense of agent outputs. F10 constructed a matrix visualization by hand, with brain areas as rows, task epochs as columns, and hypothesis verdicts as cells, specifically to make patterns visible: ``\textit{The synthesis --- the knowledge synthesis --- happens when we can put the evidence in different perspective\ldots that is a missing step right now}.'' Other examples included F11's evidence matrix, F1's optimization dashboards, F2's dependency graphs, and F9's task graph for tracing decisions (decision-level provenance). 
These varied workarounds suggest that researchers needed different representations for different sensemaking questions. Participants also took on the work of constructing these representations themselves.

\begin{tcolorbox}[title=Challenge 3: Artifact Sensemaking] Researchers struggle to synthesize across agent outputs. Reading individual artifacts does not readily reveal patterns, contradictions, and relationships across the full artifact set. Different sensemaking questions call for representations that expose different properties and relationships.

\textbf{Design implication:} Support cross-artifact synthesis through multiple coordinated representations of the same artifact set, allowing researchers to change representations as their questions change. \end{tcolorbox}

\subsubsection{\textbf{Summary and Design Implications}}

Together, these challenges suggest that post-run sensemaking requires researchers to connect three questions: what the run produced, which decisions shaped those outputs, and how the work served the research goal. Participants’ accounts revealed difficulty both answering these questions and tracing connections among them. Section~\ref{sec:framework} develops a representation framework to address these needs.

\section{A Framework for Sensemaking over Long-Horizon Agent Runs}
\label{sec:framework}

\begin{displayquote} 
``\textit{Sensemaking is the process of searching for a representation and encoding data in that representation to answer task-specific questions}.''~\cite{russell1993cost} 
\end{displayquote}

Russell et al.'s definition~\cite{russell1993cost} emphasizes the work of finding a representation and organization information within it to answer a question. For example, a researcher comparing two hypotheses might place them in table rows, with columns for their assumptions, supporting evidence, and limitations.
Their learning-loop account describes how people construct a representation (e.g., table, graph, cluster), organize available information within it, revises it when some information does not fit, and shifts to another representation until the question becomes sufficiently easier to answer.
Following Kirsh~\cite{kirsh2010thinking}, different external representations make different inferences easier or harder. A transcript, for example, supports sequential reading, while a a table can make comparison easier by aligning information along shared dimensions. Researchers therefore need representations matching the inference they are trying to make, together with sufficiently low costs of shifting among them.

\begin{figure}
    \centering
    \includegraphics[width=\linewidth, trim={7cm 1.5cm 5cm 1.5cm}, clip]{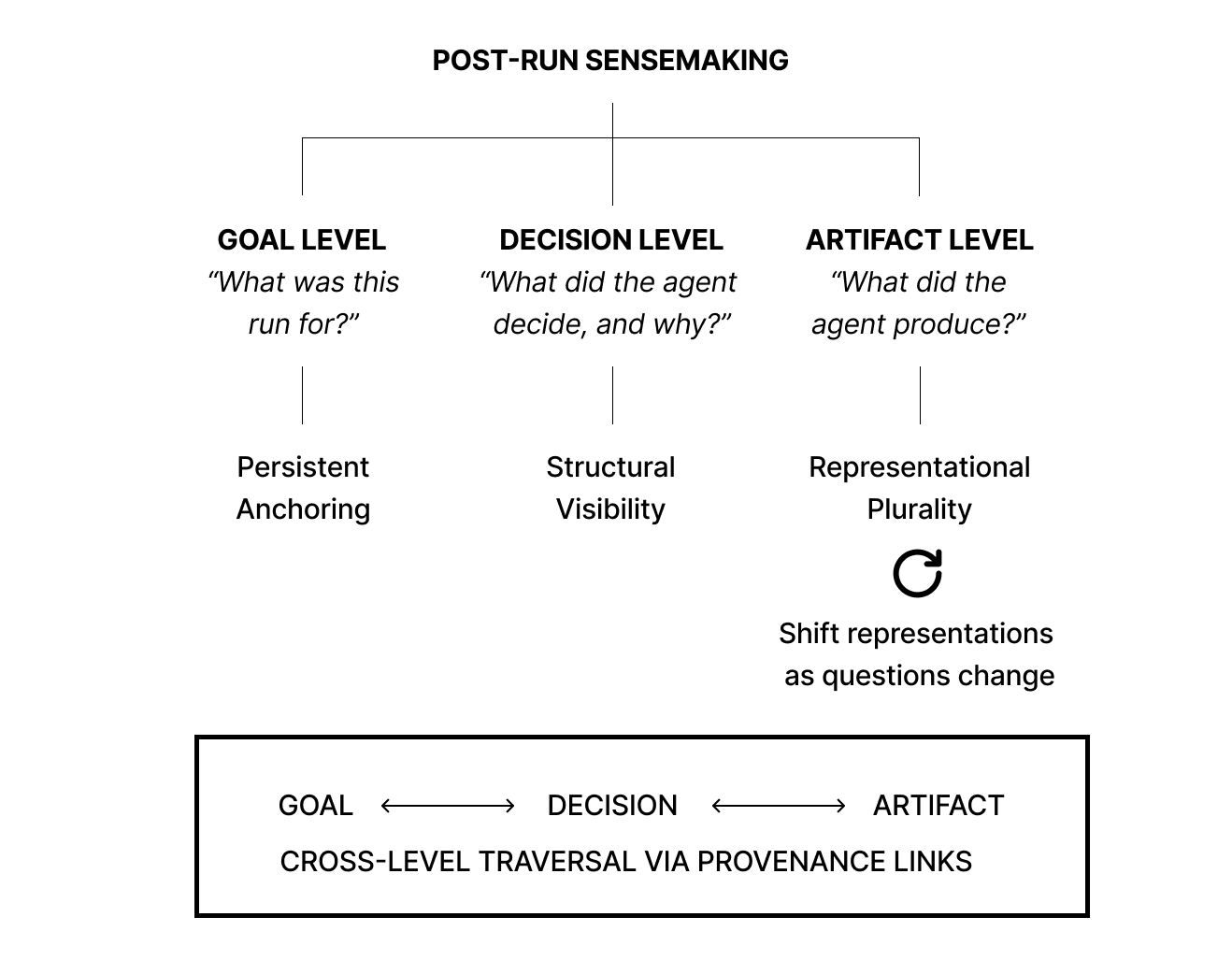}
    \caption{A representation framework for post-run sensemaking over long-horizon agent runs. The framework distinguishes three linked levels --- goal, decision, and artifact --- and a representational requirement for each. Representational shifts support different questions about the same artifact set. Provenance links enable traversal among artifacts, the decisions that shaped them, and the research goals those decisions were intended to serve.}
    \label{fig:framework}
\end{figure}

Our formative study suggests that post-run sensemaking involves this representational work at \textbf{three linked levels of a long-horizon agent run} (Fig.~\ref{fig:framework}): the \textit{goal level}, concerning what the run was for; the \textit{decision level}, concerning what the agent chose, considered, and set aside; and the \textit{artifact level}, concerning what the agent produced. Each level poses a different sensemaking question and motivates a corresponding representational requirement: \textbf{persistent anchoring} keeps the research goal accessible, \textbf{structural visibility} exposes decisions and their relationships, and \textbf{representational plurality} supports examining artifacts through different representations.

The framework distinguishes two complementary interactions that Atlas supports.
\textbf{Representational shift} changes how the same artifact set is organized or encoded to support different questions. For example, a researcher might move from a cluster view showing topic coverage to a graph showing relationships between findings and hypotheses. \textbf{Cross-level traversal}, on the other hand, moves between levels of the run, such as tracing a finding to the decision that shaped it and then evaluating that decision against the research goal. We first examine the linear execution logs as a default representation of agent activity, then derive the representational requirement associated with each level.

\subsection{Current Approach: Linear Execution Logs as the Default Representation}
\label{sec:framework-transcript}

Agents often present their actions, tool outputs, and recorded reasoning as a linear execution log, sometimes accompanied by a summary. This format supports reviewing a run in sequence but can make other sensemaking tasks more demanding. Comparing findings may require navigating between distant entries and remembering their details. Connecting outputs to the research goal may require revisiting the initial instructions and tracing subsequent actions, while recovering abandoned directions may require searching intermediate entries because those directions are not reflected in the final artifacts.

The difficulty goes beyond the length of the execution log. Its sequential structure may not match the researcher’s sensemaking question. Participants’ custom matrices, dependency graphs, and task graphs illustrate how researchers had to construct alternative representations to support sensemaking questions the execution log did not readily answer.


\subsection{Goal Level: Persistent Anchoring}
\label{sec:system-dg1}

At the goal level, researchers need to evaluate how the agent’s findings and decisions advanced what they were trying to investigate. Following an initial plan is only part of this evaluation; researchers must also assess whether the work remained relevant to the research goal. 

When the research goal is difficult to recover, researchers must reconstruct the context needed to evaluate artifacts and decisions.
Weick~\cite{weick1995sensemaking} describes sensemaking as retrospective and ongoing, with people interpreting experience in relation their projects and concerns. Klein et al.'s~\cite{klein2006sensemaking} data/frame theory describes how people interpret data through a frame.
In our setting, the research goal provides a frame for interpreting the agent's findings and decisions. 

Persistent anchoring keeps the research goal accessible throughout sensemaking and preserves traceable connections to artifacts and decisions. These connections help researchers evaluate how the agent's work served what they were trying to investigate.

\subsection{Decision Level: Structural Visibility}
\label{sec:system-dg2}

At the decision level, researchers need to reconstruct the choices that shaped the run --- which research directions the agent pursued or set aside, which assumptions and constraints it adopted, and why. A linear execution log presents these choices in sequence, leaving researchers to reconstruct their relationships and consequences.

Scaife and Rogers~\cite{scaife1996external} argue that external representations reduce cognitive load by performing inferential work that would otherwise need to be done mentally. For agent decisions, this means making relationships among choices explicit. A branching representation can distinguish pursued, rejected, and deferred directions, while links to downstream artifacts can show where implementation choices affected the results. 
Researchers can then examine whether a line of inquiry was appropriately closed, whether an alternative warrants revisiting, or whether an assumption should be reconsidered.


This imposes a representational requirement of structural visibility: the agent’s decision-making should be represented in a form that preserves consequential choices, alternatives, and their relationships. It should make visible what the agent chose, what it considered and set aside, why, and what artifacts resulted from those choices.

\subsection{Artifact Level: Representational Plurality}
\label{sec:system-dg3}

At the artifact level, researchers need to synthesize across what the agent produced. Long-horizon runs may return heterogeneous collections of papers, hypotheses, experiments, analyses, findings, and syntheses. Individual artifacts can be inspected separately, but many sensemaking questions concern patterns and relationships across the collection, such as which findings disagree, what areas of a literature were emphasized or neglected, or how artifacts relate through provenance.

Our formative study participants described constructing matrices to examine hypothesis-evidence patterns, dependency graphs to verify implementation, and task graphs to trace provenance. These examples suggest a need for representations suited to different questions. Viewed through Russell et al.’s account, participants were doing the work of constructing and adapting representations themselves.


Ainsworth's DeFT framework~\cite{ainsworth2006deft} helps explain how complementary representations suppport different operations. Clustering artifacts by topic can show coverage, aligning findings along shared dimensions in a table can support comparison, and a graph of provenance links can expose dependencies. These representations support different inferences, and using an ill-suited representation can make a question unnecessarily costly to answer~\cite{kirsh2010thinking}.

When researchers' questions change, or the current representation obscures relevant information, they should be able to shift representations without rebuilding them from scratch. 
Importantly, these representations should coordinate selection and focus so that researchers can retain context across shifts.
Otherwise, the effort of reorientation may be costly and discourage switching to a more suitable representation.

This imposes a representational requirement of coordinated representational plurality: researchers should be able to inspect the same artifact set through multiple representations suited to different sensemaking questions while retaining context across shifts.

\subsection{From Representational Requirements to Design Goals}

We translate these representational requirements into three design goals (\textbf{DGs}) for systems supporting post-run sensemaking.

\begin{itemize}
    \item \textbf{DG1: Anchor the Research Goal and Enable Provenance Traversal}. Keep the research goal as a reference point and link it to decisions and artifacts, so researchers can evaluate whether what the agent found and decided is relevant and productive for what they are trying to investigate.
    \item \textbf{DG2: Make Agent Decisions Structurally Visible}. Represent the agent’s decision-making in a structure that preserves alternatives alongside the pursued path. Surface what the agent chose, what it considered and set aside, why, and what resulted from those decisions so researchers can assess how decisions shaped the run.
    \item \textbf{DG3: Support Cross-Artifact Sensemaking Through Coordinated Representations}. Provide multiple representations of the same artifact set for different sensemaking questions. Keep artifacts recognizable across representations and preserve selection and focus so researchers can switch views without losing their place.
\end{itemize}

These design goals specify what a workspace should support for post-run sensemaking without prescribing a particular interface. Atlas implements the framework through a unified canvas that keeps the research goal accessible, exposes the agent's decision trajectory, and presents the same artifact set through multiple coordinated representations. Provenance links connect goals, decisions, and artifacts, allowing researchers to trace how decisions shaped the outputs and how the work related to the research goal. The next section describes how Atlas provides this support.


\section{System: \sys}
\label{sec:system}

Atlas implements these requirements through a unified canvas that links the agent's trajectory to its artifacts. Researchers can inspect the artifacts associated with a trajectory node and trace an artifact back to the decisions that shaped it. The canvas supports two complmenetary modes, artifact projections for cross-artifact sensemaking and a trajectory tree for inspecting decisions. Links between them allow researchers to move from examining outputs to investigating how they arose.


\subsection{Artifact Projections (DG3)}
\label{sec:artifact-projections}

\begin{figure*}[h!]
    \centering
    \includegraphics[width=\linewidth]{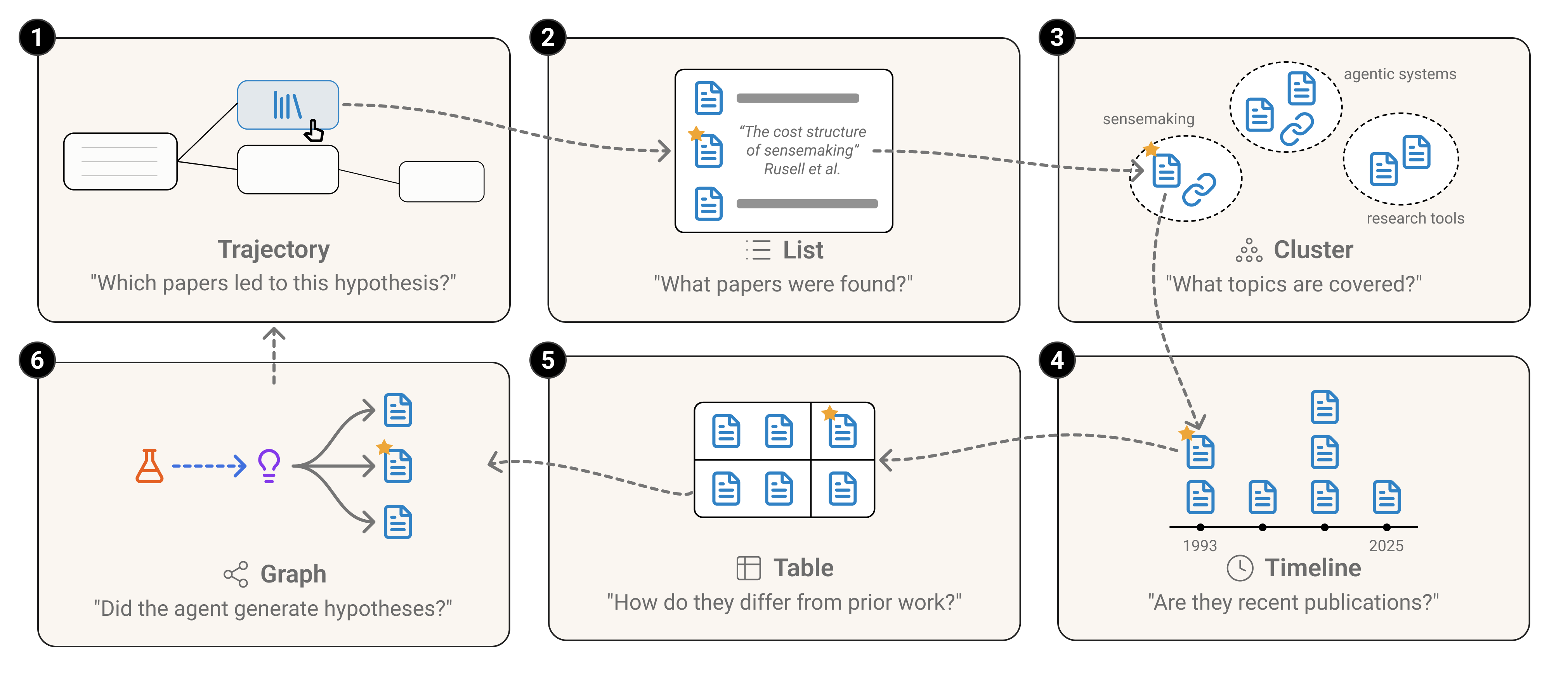}
    \caption{\textbf{Walkthrough of a researcher’s exploration as their sensemaking questions change.} (1) ``\textit{Where do these hypotheses come from?}'' The researcher traces a hypothesis to the papers that informed it. (2) ``\textit{What topics do these papers cover?}'' After noticing the \textit{Cost Structure of Sensemaking} paper, the researcher explores the surrounding literature, first switching to the \raisebox{-2pt}{\includegraphics[scale=0.4]{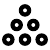}} \textbf{cluster} projection to see the subtopics covered by the paper set and then to (3) the \raisebox{-2pt}{\includegraphics[scale=0.4]{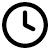}} \textbf{timeline} to inspect when the papers were published.  (4) ``\textit{How are these papers different?}'' The researcher switches to the \raisebox{-2pt}{\includegraphics[scale=0.4]{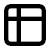}} \textbf{table} projection to compare the papers across dimensions. (5) ``\textit{Are there any hypotheses generated from these papers?}'' Finally, they switch to \raisebox{-2pt}{\includegraphics[scale=0.4]{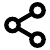}} \textbf{graph} to check for any artifacts that are linked to these papers and then return to the node in trajectory from where they were fetched. The sequence illustrates how Atlas supports movement across trajectory and artifact representations as the researcher’s question changes. (See Fig.~\ref{fig:example-screenshot} for the corresponding interactions in the implemented Atlas interface.)}
    
    \label{fig:projection-scenario}
\end{figure*}

The canvas displays \raisebox{-2pt}{\includegraphics[scale=0.4]{figures/icons/artifact-paper.pdf}} papers, \raisebox{-2pt}{\includegraphics[scale=0.4]{figures/icons/artifact-hypothesis.pdf}} hypotheses, \raisebox{-2pt}{\includegraphics[scale=0.4]{figures/icons/artifact-experiment.pdf}} experiments, \raisebox{-2pt}{\includegraphics[scale=0.4]{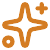}} findings, \raisebox{-2pt}{\includegraphics[scale=0.4]{figures/icons/artifact-synthesis.pdf}} syntheses as persistent objects with metadata and typed relationships to other artifacts. Artifacts remain available across sessions and retain the same identifier across projections. References in chat resolve to these same objects, connecting the conversation to the artifacts displayed on the canvas.


Researchers can view the same artifact set through six projections (Table~\ref{tab:projections}). Selection, focus, and bookmarks carry across projections, so the researcher keeps their place when switching between them. Figure~\ref{fig:projection-scenario} illustrates how the same \raisebox{-2pt}{\includegraphics[scale=0.4]{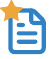}} paper appears in different positions as the representation changes.

\begin{table*}[t]
\caption{The six artifact projections, the sensemaking operations they support, and example questions.}
\label{tab:projections}
\small
\resizebox{0.9\linewidth}{!}{
\begin{tabular}{@{}p{1.7cm} p{5.8cm} p{6.5cm}@{}}
\toprule
\textbf{Projection} & \textbf{Affordance} & \textbf{Example question} \\
\midrule

\raisebox{-2pt}{\includegraphics[scale=0.4]{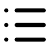}}~\textbf{List}
& \textbf{Inventory and item-level inspection}. \newline Shows which artifacts are present and supports sorting by artifact metadata.
& 
``What artifacts did the run collect or produce?'' \newline ``Which papers are most relevant or recent?''\\

\addlinespace

\raisebox{-2pt}{\includegraphics[scale=0.4]{figures/icons/layout-timeline-icon.pdf}}~\textbf{Timeline}
& \textbf{Temporal ordering}. \newline Places artifacts along an appropriate temporal dimension, such as publication date for papers or production time for run artifacts.
& 
``When were these papers published?''\newline  ``At what point in the run did these findings emerge?''\\

\addlinespace

\raisebox{-2pt}{\includegraphics[scale=0.4]{figures/icons/layout-cluster-icon.pdf}}~\textbf{Cluster}
& \textbf{Thematic grouping and coverage}.\newline  Groups artifacts by similarity to reveal themes, concentrations, gaps, and isolated items.
& 
\makecell[l]{``Which topics are represented in this collection?'' \\
``Which topics have many artifacts, and which have few?''}
\\

\addlinespace

\raisebox{-2pt}{\includegraphics[scale=0.4]{figures/icons/layout-table-icon.pdf}}~\textbf{Table}
& 
\textbf{Structured comparison across dimensions}. Organizes artifacts into rows and columns defined by two dimensions, allowing researchers to compare similarities, differences, and combinations.
& 
``How do they differ in methodology and application?''\newline ``Which methods have been used for which application areas?''\\

\addlinespace

\raisebox{-2pt}{\includegraphics[scale=0.4]{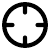}}~\textbf{Cartesian}
& 
\textbf{Relative position, tradeoffs, and outliers}.\newline Places artifacts along two continuous dimensions to expose unusual combinations and extremes.

& 
``Which hypotheses are both highly novel and feasible?'' \newline ``Which artifact is an outlier relative to the rest?'' \\

\addlinespace

\raisebox{-2pt}{\includegraphics[scale=0.4]{figures/icons/layout-graph-icon.pdf}}~\textbf{Graph}
& 
\textbf{Relational and provenance structure}. \newline Makes typed relationships between artifacts explicit and supports tracing how one artifact grounds, tests, supports, or contradicts another.
& 
``Which papers ground this hypothesis?''\newline ``Which experiment tested it, and what findings did it produce?'' \\

\bottomrule
\end{tabular}
}
\end{table*}

\begin{figure*}[h!]
  \centering
  \includegraphics[width=0.8\textwidth]{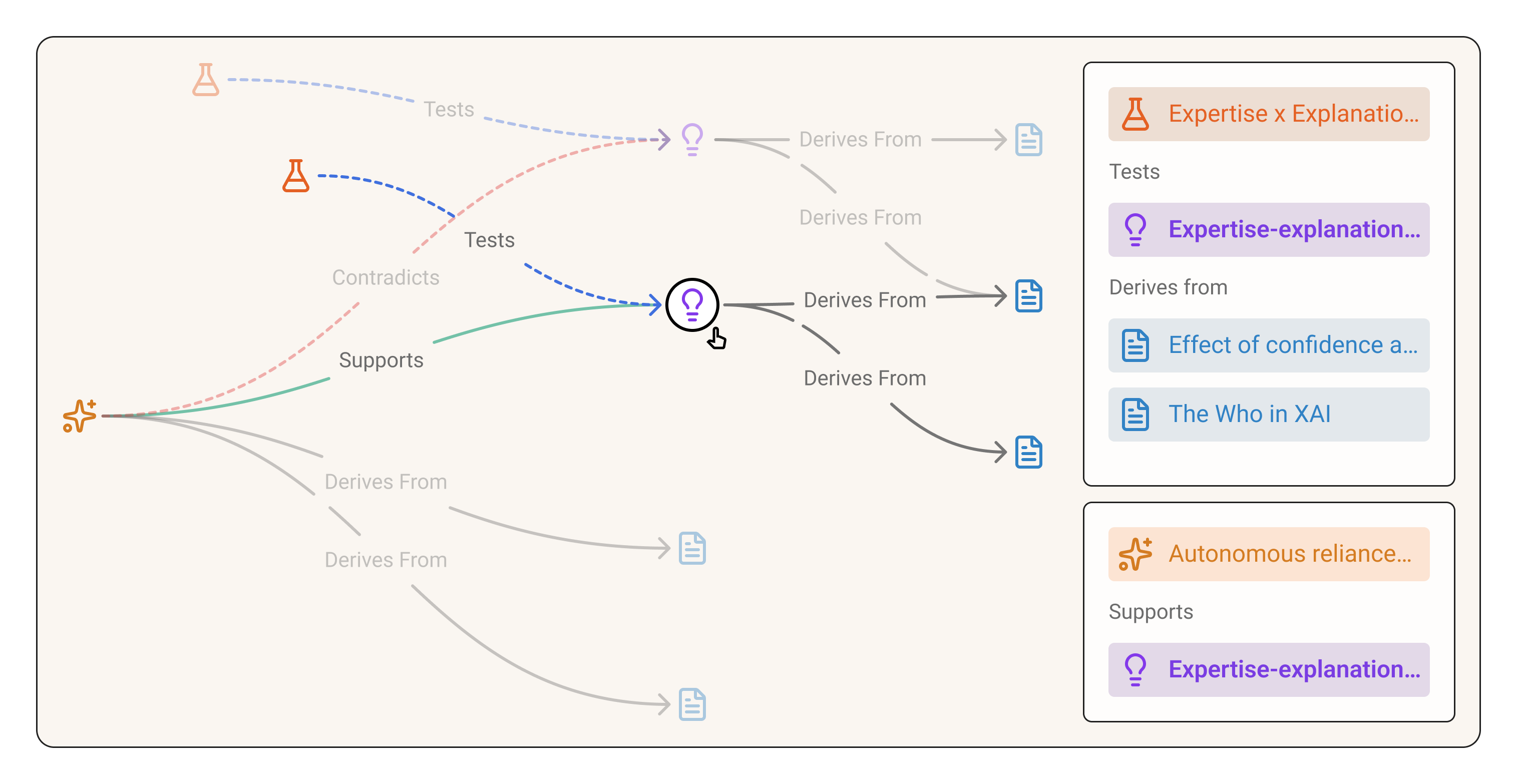}
\caption{\textbf{The graph projection.} It shows the relations between artifacts, if any. For example, it may show relationships between papers and hypothesis ({\raisebox{-2pt}{\includegraphics[scale=0.4]{figures/icons/artifact-hypothesis.pdf}}} hypothesis \texttt{derives\_from} {\raisebox{-2pt}{\includegraphics[scale=0.4]{figures/icons/artifact-paper.pdf}}} paper), experiment and hypothesis ({\raisebox{-2pt}{\includegraphics[scale=0.4]{figures/icons/artifact-experiment.pdf}}} experiment \texttt{tests} {\raisebox{-2pt}{\includegraphics[scale=0.4]{figures/icons/artifact-hypothesis.pdf}}} hypothesis), finding and paper ({\raisebox{-2pt}{\includegraphics[scale=0.4]{figures/icons/artifact-finding.pdf}}} finding \texttt{derives\_from} {\raisebox{-2pt}{\includegraphics[scale=0.4]{figures/icons/artifact-paper.pdf}}} paper), and finding and hypothesis ({\raisebox{-2pt}{\includegraphics[scale=0.4]{figures/icons/artifact-finding.pdf}}} finding \texttt{supports} or \texttt{contradicts} {\raisebox{-2pt}{\includegraphics[scale=0.4]{figures/icons/artifact-hypothesis.pdf}}} hypothesis).}
  \label{fig:layout-graph-provenance}
\end{figure*}

Five projections organize artifacts by their attributes, while the graph displays typed relationships among them. Researchers can also follow provenance links from artifacts to the trajectory tree to inspect the decisions that produced them. 

The axes available in the Cartesian projection depend on the artifact type. For papers, researchers can select from a dropdown menu of candidate dimensions derived from the corpus or use a text field to ask the AI to generate dimensions for their questions. Hypotheses are positioned by feasibility and novelty, while findings are positioned by confidence and novelty.

\subsection{Trajectory Tree (DG1, DG2)}
\label{sec:trajectory-tree}

\begin{figure*}[h!]
  \centering
  \includegraphics[width=\textwidth]{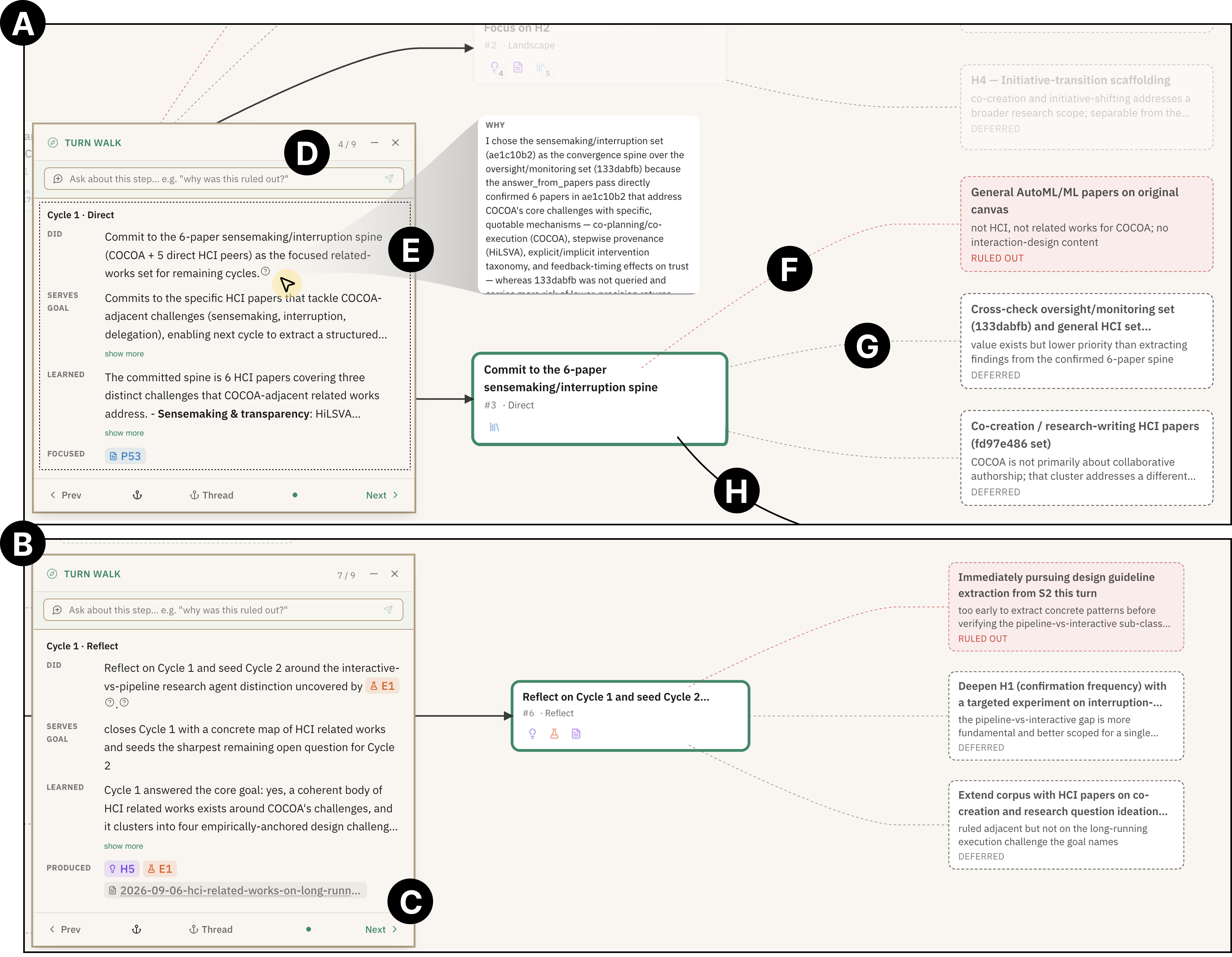}
  \caption{\textbf{Trajectory walkthrough} at (A) the fourth and (B) seventh steps of a run. Users can navigate sequentially using (C) Previous and Next, (D) ask questions about the current step, and (E) inspect its node card. The card summarizes what the agent did and why, how the step relates to the research goal, what it learned, and which artifacts it used or produced. Branches show directions the agent (F) ruled out, (G) deferred, or (H) pursued. Selecting a node reveals its recorded rationale.}
  \label{fig:trajectory}
\end{figure*}

\begin{figure*}[h!]
  \centering
  \includegraphics[width=\textwidth]{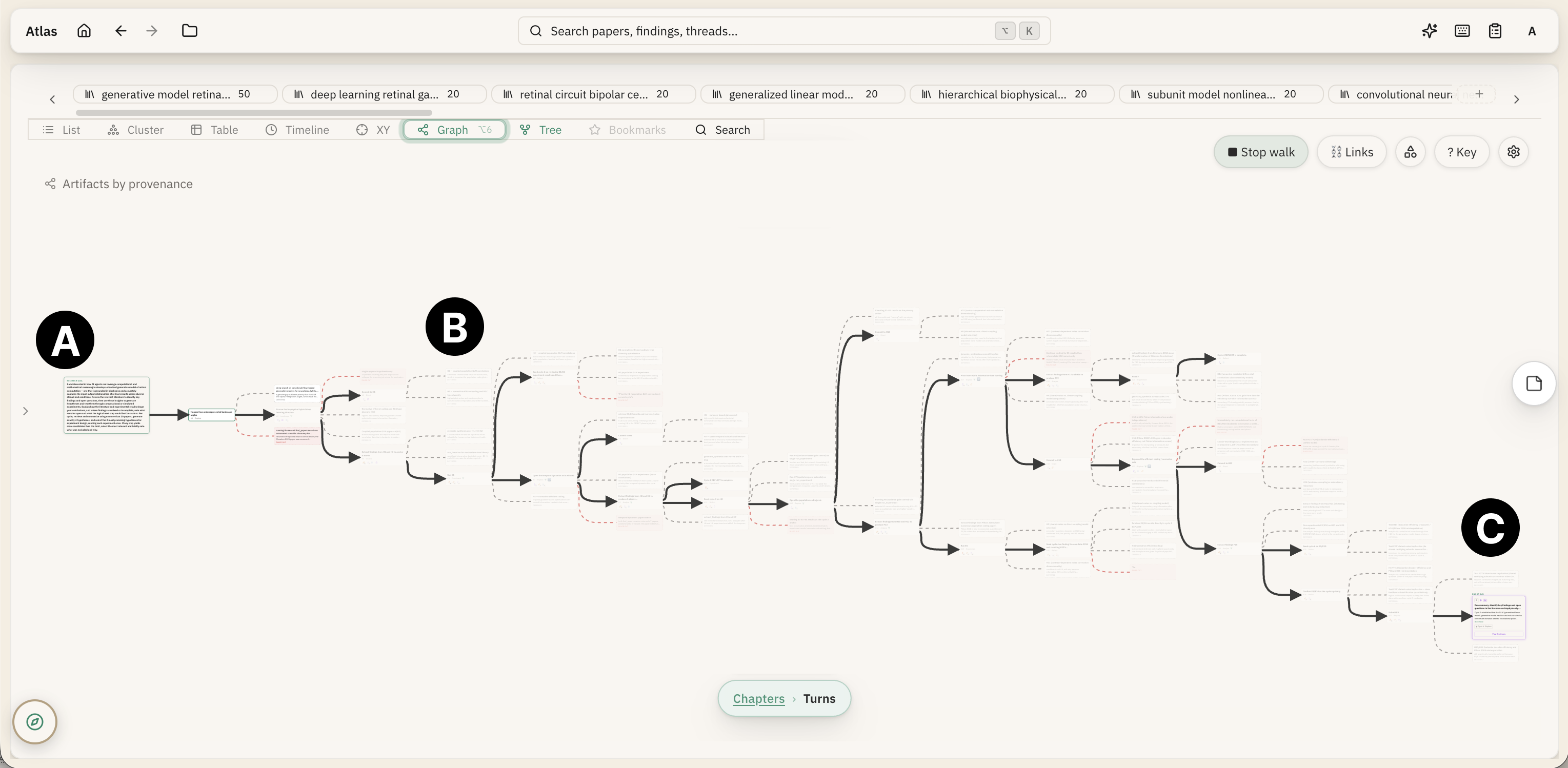}
\caption{\textbf{Trajectory of a 1.5-hour \textsc{Extended Run}} P2 reviewed in the user study. The run retrieved \raisebox{-2pt}{\includegraphics[scale=0.4]{figures/icons/artifact-paper.pdf}} 207 papers, and generated \raisebox{-2pt}{\includegraphics[scale=0.4]{figures/icons/artifact-hypothesis.pdf}} 24 hypotheses, \raisebox{-2pt}{\includegraphics[scale=0.4]{figures/icons/artifact-finding.pdf}} 29 findings, \raisebox{-2pt}{\includegraphics[scale=0.4]{figures/icons/artifact-synthesis.pdf}} 2 syntheses, and \raisebox{-2pt}{\includegraphics[scale=0.4]{figures/icons/artifact-experiment.pdf}} 10 experiments. The trajectory included 49 deferred and 10 ruled-out options and reached a maximum depth of 15 levels. (A) The research goal anchors the root. (B) Agent turns form nodes, with branches showing pursued and alternative directions. (C) The trajectory ends with a summary of the completed run.}
  \label{fig:trajectory-overview}
\end{figure*}

The trajectory tree makes the agent's decisions inspectable and places the research goal at its root (Fig.~\ref{fig:trajectory-overview}). Each subsequent node represents one agent turn, with branches showing where the agent pursued different directions. Researchers can follow links from artifacts to associated turns and trace the trajectory back to the research goal.

\textbf{Node cards (Fig.~\ref{fig:trajectory}E).} Clicking a node opens a card showing what the agent did, the artifacts associated with that step, and how the step was intended to advance the research goal. The goal remains visible at the root, while each card explains the contribution of an individual step. Artifacts appear inline and can also be inspected through the artifact projections.

\textbf{Branching structure: pursued, deferred, and ruled-out paths (Fig.~\ref{fig:trajectory}F–H).} 
Branches show the paths the agent pursued, deferred, or ruled-out during the run. The alternative paths the agent deferred or ruled out remain visible alongside pursued paths, making them easier to locate than when embedded in a linear execution log. Ruled-out paths include a reason for closure, while deferred paths remain open for later exploration. In the chapter-level overview, sibling ruled-out paths are grouped into a single tile to keep the tree readable. Zooming in reveals the individual paths and the reasons recorded for each.

\textbf{Walk mode (Fig.~\ref{fig:trajectory}).} A walk panel presents the run in turn order, allowing researchers to review the agent's sequence without navigating the tree spatially. Switching between walk mode and the tree mode preserves the current turn.

\subsection{Usage Scenario}
\label{sec:usage-scenario}

Paul is an NLP researcher studying retrieval-augmented generation (RAG). Before leaving workf or the day, he submits the research goal, ``\textit{I am interested in understanding when and why retrieval-augmented generation fails on tasks that require reasoning across long contexts}'', and asks Atlas to investigate it autonomously.

\textbf{What the agent did.} The agent runs for 1.5 hours in \textsc{Extended Run} mode and returns 156 artifacts: \raisebox{-2pt}{\includegraphics[scale=0.4]{figures/icons/artifact-paper.pdf}} 120 papers, \raisebox{-2pt}{\includegraphics[scale=0.4]{figures/icons/artifact-hypothesis.pdf}} 14 hypotheses, \raisebox{-2pt}{\includegraphics[scale=0.4]{figures/icons/artifact-synthesis.pdf}} 6 syntheses, \raisebox{-2pt}{\includegraphics[scale=0.4]{figures/icons/artifact-experiment.pdf}} 6 experiments, and \raisebox{-2pt}{\includegraphics[scale=0.4]{figures/icons/artifact-finding.pdf}} 10 findings. Paul opens Atlas the next morning.

\textbf{Surveying the landscape.} The canvas opens in list view, artifacts grouped by the cycle that produced them. Paul switches to the \raisebox{-2pt}{\includegraphics[scale=0.4]{figures/icons/layout-cluster-icon.pdf}} cluster projection to get a sense of the territory the agent covered. Three clusters emerge: work on retrieval precision and noise, work on context window limitations in transformer architectures, and a smaller cluster around multi-hop reasoning benchmarks. Paul notices the third cluster contains fewer papers and makes a note that the agent may not have explored it deeply.

\textbf{Seeing what the agent pursued and what it set aside.} Paul switches to the trajectory tree (Fig.~\ref{fig:trajectory},~\ref{fig:trajectory-overview}). 
During the exploration phase, the agent identified six candidate research directions and pursued three: retrieval noise under long-context conditions, position bias in retrieved passages, and the effect of chunk size on reasoning coherence. Three other candidate directions --- including the multi-hop reasoning thread Paul noticed in the cluster view --- appear as ruled-out siblings, each with a recorded reason. 
The recorded reason states that the agent ruled out further investigation of multi-hop reasoning because the question was underspecified given the available literature. This helps Paul understand why the corresponding cluster contains fewer papers.
Paul flags this as a direction worth revisiting.

\textbf{Tracing a finding to its supporting evidence.} One finding catches Paul's attention: the agent reports that position bias in retrieved passages accounts for a larger performance drop than retrieval noise under long-context conditions, and rates it high confidence. 
He wants to understand the evidence behind this finding. Paul follows its provenance link to the associated hypothesis in the trajectory tree and opens the node card.
The card records that the agent generated the hypothesis after synthesizing four papers on \textbf{\hlcb{paperColor}{\raisebox{-2pt}{\includegraphics[scale=0.4]{figures/icons/artifact-paper-set.pdf}}attention sink phenomena}} and designed an experiment to test it by varying passage position in a fixed retrieval set across three benchmark tasks. 

Paul opens the experiment in full viewer mode, where he inspects the code and results table. He notices that the agent used exact-match accuracy as the evaluation metric, which he considers too restrictive for the open-ended reasoning tasks. He now questions whether the reported performance difference reflects reasoning quality or sensitivity to answer wording.

\textbf{Connecting back to the goal.} Paul switches back to the \raisebox{-2pt}{\includegraphics[scale=0.4]{figures/icons/layout-cartesian-icon.pdf}} Cartesian projection and compare findings by confidence and novelty. The position-bias finding appears in the high-confidence, high-novelty quadrant alongside a contradictory finding from another experiment that used a generative evaluation metric and reported a smaller effect. Paul returns to the trajectory tree, where the experiments appear as sibling nodes, and opens their cards to compare the recorded design rationale. He opens the experiments in full viewer mode to inspect their evaluation methods and results. In relation to his original question about when and why RAG fails, he treats position bias as a direction worth investigating but considers the magnitude of its effect unresolved.

Paul moves among artifact projections, the trajectory tree, node cards, and experiment viewers as his questions change. These linked views help identify an underexplored direction, inspect the evaluation choices behind a finding, and assess what the run contributes to his research goal.

\subsection{Implementation}
\label{sec:implementation}

Atlas is a web application built with React. The canvas is drawn directly, without a diagramming library. A d3 force simulation positions artifacts in the cluster projection. Provenance graphs use Dagre for layout, switching to Elk for graphs with more than 80 nodes to improve edge routing. Agent turns run through the Claude Agent SDK over a surface of 71 tools, using Claude Sonnet 4.6 for the agent and Haiku 4.5 for the short auxiliary passes that run on every turn. Paper search uses the Semantic Scholar API, and experiments execute as Python in an isolated container. The agent runs in one of two modes: (1) \textsc{Auto}, which offers a typical chat (back-and-forth) interaction experience, and (2) \textsc{Extended Run}, in which it works unattended for a fixed time budget or until a specified goal is satisfied. In \textsc{Extended Run}, the agent works through repeated research cycles consisting of five phases: \textsf{Explore}, \textsf{Direct}, \textsf{Analyze}, \textsf{Experiment}, and \textsf{Reflect}. The cycles continue until the agent determines that the specified goal has been satisfied or the specified time budget is reached. We describe the execution procedure, phase prompts, and trajectory reconstruction in Appendix~\ref{app:extended-run}.

\section{User Study}
\label{sec:user-study}

\subsection{Procedure}
We conducted a within-subjects study with 14 participants to examine how Atlas supports post-run sensemaking compared with a standard chat-and-artifact workspace. Before the study session, each participant submitted two research goals. We generated one agent run for each goal using Atlas in \textsc{Extended Run} mode with the same underlying generation pipeline, phase prompts, and tools. Participants reviewed one run in each condition, with system order counterbalanced across participants. Across the 28 study runs (= 14 participants x 2 topics), they contained an average of 170.4 artifacts (\(SD=78.3\)). Trajectories contained an average of 15.9 turns, 22.2 deferred options, and 6.2 ruled-out options per run. Artifact counts varied substantially by topic, primarily because of variation in literature retrieval, while counts of higher-level research artifacts were more similar across conditions. Agent execution lasted 83.9 minutes on average (\(SD=23.7\)). Appendix~\ref{app:run-statistics} reports the full run-level descriptive statistics and paired comparisons.

\subsection{Research Questions}

We examine how Atlas supports researchers’ sensemaking of completed agent runs, along with the subjective workload and difficulty of maintaining orientation during that process.


\begin{description}[leftmargin=0pt,itemsep=4pt]
  \item \textbf{RQ1: Trajectory Sensemaking}. How does Atlas support researchers in understanding consequential agent decisions, tracing their relationship to the research goal, and identifying productive next steps compared with the baseline workspace?

  \item \textbf{RQ2: Artifact Sensemaking}. How does Atlas support researchers in synthesizing patterns and relationships across a run’s artifacts compared with the baseline workspace?

  \item \textbf{RQ3: Review Workload and Orientation}. How does reviewing a completed run in Atlas compare with the baseline in subjective workload and researchers’ ability to keep track of their place and focus?
\end{description}

The review tasks were derived from the problems participants described in the formative study, rather than from the specific interaction techniques implemented in Atlas. Participants could use any available functionality, including asking the agent for summaries or structured outputs in the baseline chat panel 
(e.g., see visual outputs in chat in Fig.~\ref{fig:baseline-visual-in-chat}).

\subsection{Conditions}

\begin{figure*}
    \centering
    \includegraphics[width=\linewidth]{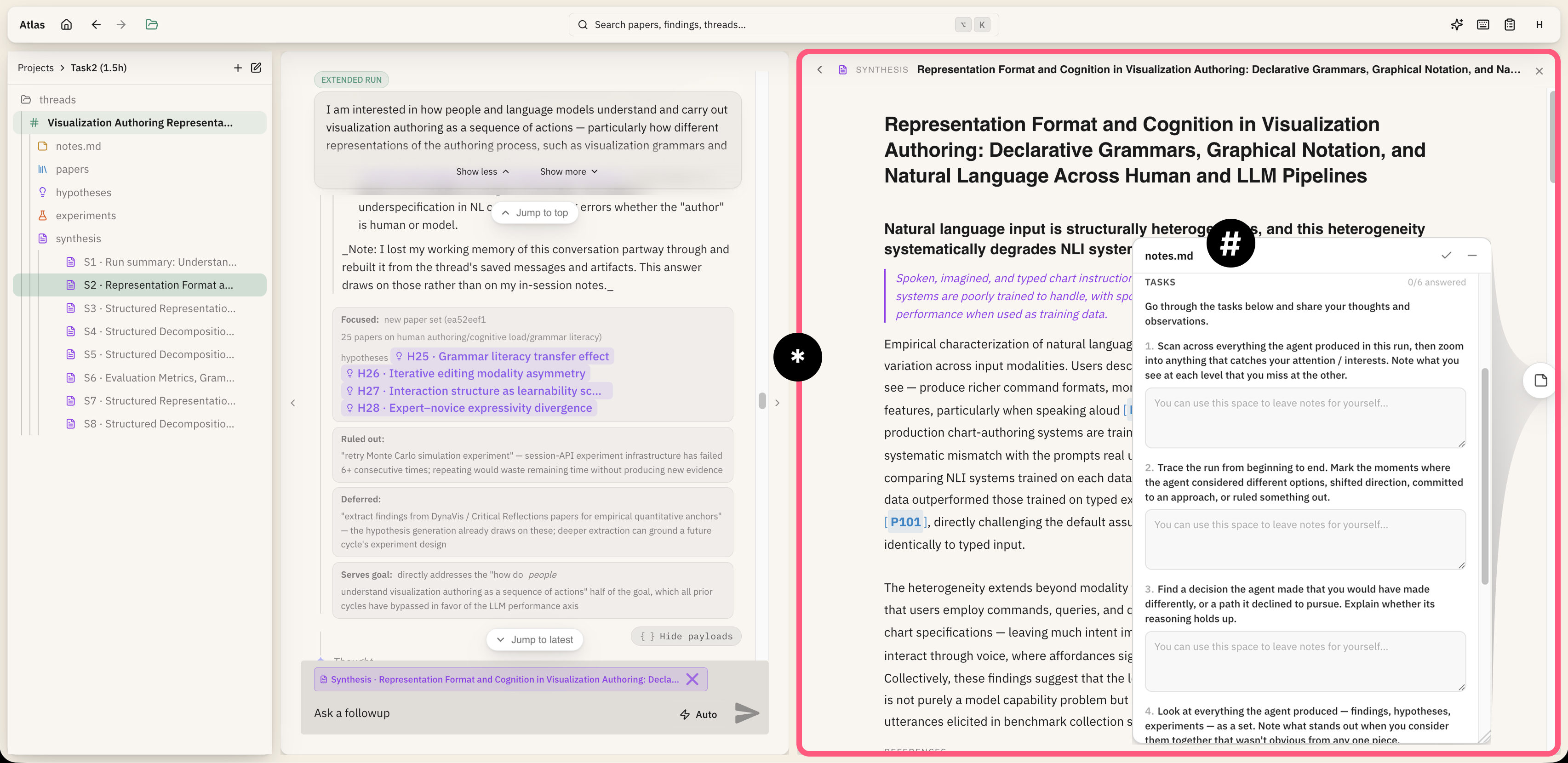}
    \caption{\textbf{The baseline interface pairs a chat panel (center) with an artifact viewer (right), shown displaying a \raisebox{-2pt}{\includegraphics[scale=0.4]{figures/icons/artifact-synthesis.pdf}} synthesis report the run produced. The run's narration, including pursued, ruled-out, and deferred directions, appears in the chat panel. The highlighted viewer region (*) is the only region that differs between conditions; Atlas displays its canvas there. The notes viewer (\#) presents the six review-task prompts used in both conditions. See Table~\ref{tab:tasks} for the full prompts.}}
    \label{fig:baseline-interface}
\end{figure*}

\textbf{Standard workspace (Baseline; Fig.~\ref{fig:baseline-interface})}. The baseline is designed to represent the typical IDE environment that most researchers using coding agents work in today — a chat pane paired with an artifact viewer, such as a PDF viewer for papers. Artifacts are accessible and the agent can summarize, retrieve, and synthesize on request, but there is no unified canvas, no trajectory tree, and no multi-representational projection of the artifact set. They could, however, prompt the agent to produce a visually structured output in the chat panel, as they would in tools such as on Claude web app (see Fig.~\ref{fig:baseline-visual-in-chat} in Appendix).

\textbf{Atlas.} Atlas provides the baseline functionality alongside a canvas containing a trajectory tree and coordinated artifact projections. Researchers can navigate in both directions between trajectory nodes and their associated artifacts.

\subsection{Participants}
\label{sec:participants}

We recruited 14 researchers (age $M=27.3$, range 20--35) who use AI agents in their research. Eight reported 3--5 years of research experience, five reported 6--10 years, and one more than 10 years.
Participants' self-reported research domains included HCI (N=9), NLP (N=7), AI/ML (N=5), Neuroscience (N=2), and AI governance, Visualization, and Visual Analytics (N=1 each). Every participant used literature review as a method, 12 ran computational experiments, 12 ran user studies, and 9 built systems. All 14 had used research agents, and 6 had built one. Participants reported using AI agents that autonomously execute multiple research steps daily or almost daily (N = 2), a few times a week (N = 4), about once a week (N = 3), a few times a month (N = 3), or less than once a month (N = 2). Six participants reported typical agent runs lasting up to 30 minutes, one hour (N=3), and four hours or more (N=5), including eight hours or more for four participants.
Participants reported varying output volumes, from short summaries to extensive collections of documents and files. Of the 14 participants, 10 reported long documents, many outputs, or several files, and seven reported volumes too large to review completely. 

Participants commonly reviewed outputs selectively: 
13 reported skimming and prioritizing outputs that appeared important, 12 asked the agent follow-up questions, and two read a run’s outputs sequentially from beginning to end.
Seven compared the results against the original research goal, and one asked a colleague to sanity-check the work. When asked how well they typically understand \emph{why} an agent made the decisions it did, participants' median rating was 5 on a 7-point scale ($M=4.29$, \sd{1.59}; 1 not at all, 7 very well); six participants gave ratings of 3 or below. Ten participants reported feeling overwhelmed by output volume. Nine reported that making sense of a completed run was often or always mentally demanding ($M = 3.71$ on a 5-point scale). Five reported often or always feeling uncertain whether outputs addressed their research goal ($M = 3.29$). The sample therefore arrived already reporting the three difficulties this paper addresses, with respect to their own tools.

\begin{figure*}[h!]
    \centering
    \includegraphics[width=\linewidth]{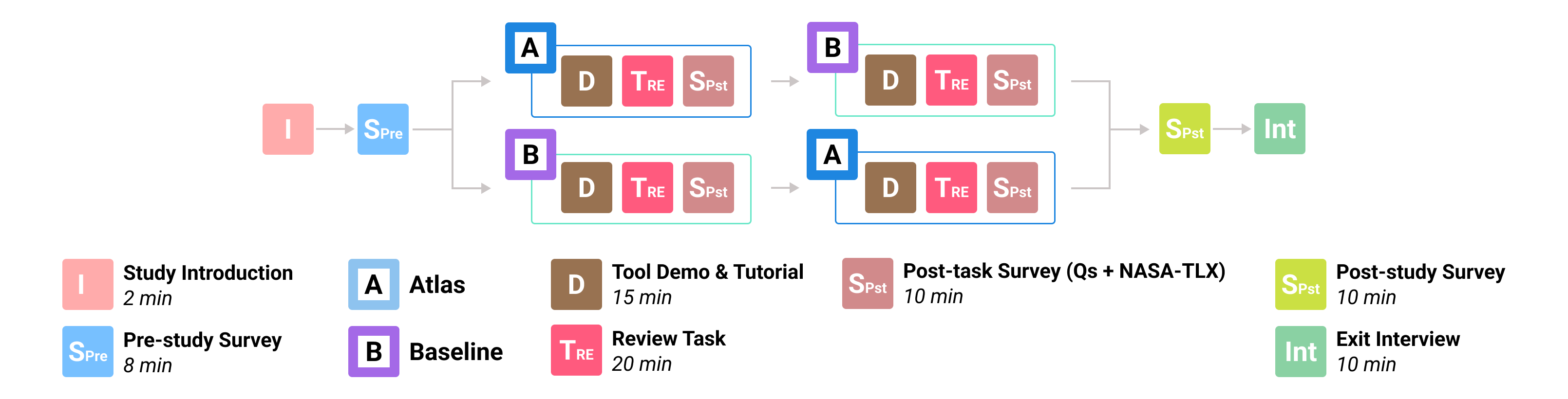}
\caption{\textbf{User study procedure.} After an introduction and a pre-study survey, each participant completed two blocks in counterbalanced order, one with Atlas (A) and one with the baseline (B). Each block consisted of a tool demo and tutorial (D, 20 min), the six-part review task (T\textsubscript{RE}, 20 min), and a post-task survey with the construct scales and NASA-TLX (S\textsubscript{Pst}, 10 min). A post-study comparative survey and an exit interview closed the session.}
    \label{fig:study-procedure}
\end{figure*}

\section{Results}
\label{sec:results}

All participants except P9 favored Atlas on the overall-preference survey item. P9 expressed that he `slightly favors' the baseline on that item, explaining that he wanted richer canvas features, such as sizing artifact icons by relevance and linking artifacts across projections placed side by side (e.g., \raisebox{-2pt}{\includegraphics[scale=0.4]{figures/icons/layout-cluster-icon.pdf}} cluster and \raisebox{-2pt}{\includegraphics[scale=0.4]{figures/icons/layout-table-icon.pdf}} table next to each other). On the direct-comparison item, however, P9 preferred Atlas and also reported lower perceived workload with it. Across the five post-task scales and on the single-item provenance measure, participants generally rated Atlas higher, and reported substantially lower subjective workload, primarily in mental demand, temporal demand, and frustration. We organize the results around trajectory sensemaking (RQ1), artifact sensemaking (RQ2), and review workload and orientation (RQ3). No interface-order effects were observed on any of the six post-task measures (all $p \geq .23$). Unless noted, comparisons between conditions are Wilcoxon signed-rank tests, Holm-corrected within each instrument.

\subsection{RQ1: Trajectory Structure Supports Decision, Provenance, and Next-Step Sensemaking}

Among the five multi-item post-task scales, decision transparency showed the largest difference between conditions. Participants rated Atlas well above the baseline (Atlas median 6.0 vs.\ baseline 3.8, $p = .0002$), and every participant preferred Atlas for understanding what the agent decided and why.

Much of this difference concerned whether participants could identify directions the agent had considered but not pursued. Importantly, the baseline execution log contained the same ruled-out, deferred, and pursued paths, as shown in Fig.~\ref{fig:baseline-interface}. Participants nevertheless often experienced these alternatives as effectively invisible because locating them required recovering consequential choices from sequential narration. The difference was therefore not information availability but legibility. Atlas externalized these decisions as persistent structure. P4 said: ``\textit{This declined-to-pursue path was much more difficult to identify in the previous [baseline] interface}''; ``\textit{one thing that I really do like about [Atlas] is how it's more transparent \ldots it also shows something that the agent thought about doing but didn't do}'' [P7]. 

Making these alternatives visible also changed how participants reasoned about what to do next. Eight participants treated these ruled-out and deferred paths as an inventory of possibilities they could revisit. P14 noted that they were ``\textit{not restricted to just the last step that the agent thought was interesting},'' because the tree exposed hypotheses and findings that had been deferred, saying ``\textit{I was able to see \ldots hypotheses and findings that were deferred, whereas I wasn't able to see that in the \ldots linear form}.''
P4 said: ``\textit{you have a sense of the possibilities, but which decision or track you chose to investigate further}.'' P5 wanted to act on that sense in place: ``\textit{then I can look at these [paths] and I could [tell] that was actually closer \ldots And I could continue from this branch}.'' P11 noticed that the agent had committed to two hypotheses out of five: ``\textit{it proposed four or five hypotheses, and it selected only two to evaluate, and it was clear to me \ldots I'm going to choose these [other] two [as my follow-up experiments]}.'' P4 immediately picked up a declined direction as their own next move: ``\textit{I would have pursued my next experiment to be a follow-up of the previous one}.'' Every participant reported greater confidence in identifying next steps with Atlas, the strongest effect in the study ($p = .0001$).

Participants rated Atlas higher for tracing findings to the decisions and supporting evidence that shaped them. For P2, this traceability affected what they felt they can do with a result: ``\textit{when I trust your provenance, I can say, oh, this link or this node is doing X, Y, or Z, so I would do this.}''

Goal connection also improved more modestly, and responses varied across participants. Five participants gave identical ratings across conditions, while every non-tied comparison favored Atlas. Several participants specifically valued being able to see how individual step was intended to contribute to the goal (Fig.~\ref{fig:trajectory}E). P2, for example, had identified before using Atlas that understanding how individual steps connected to the research goal was the most important information for evaluating a run, echoing concerns raised in our formative study.

\begin{figure*}[h!]
  \centering
  \includegraphics[width=\textwidth]{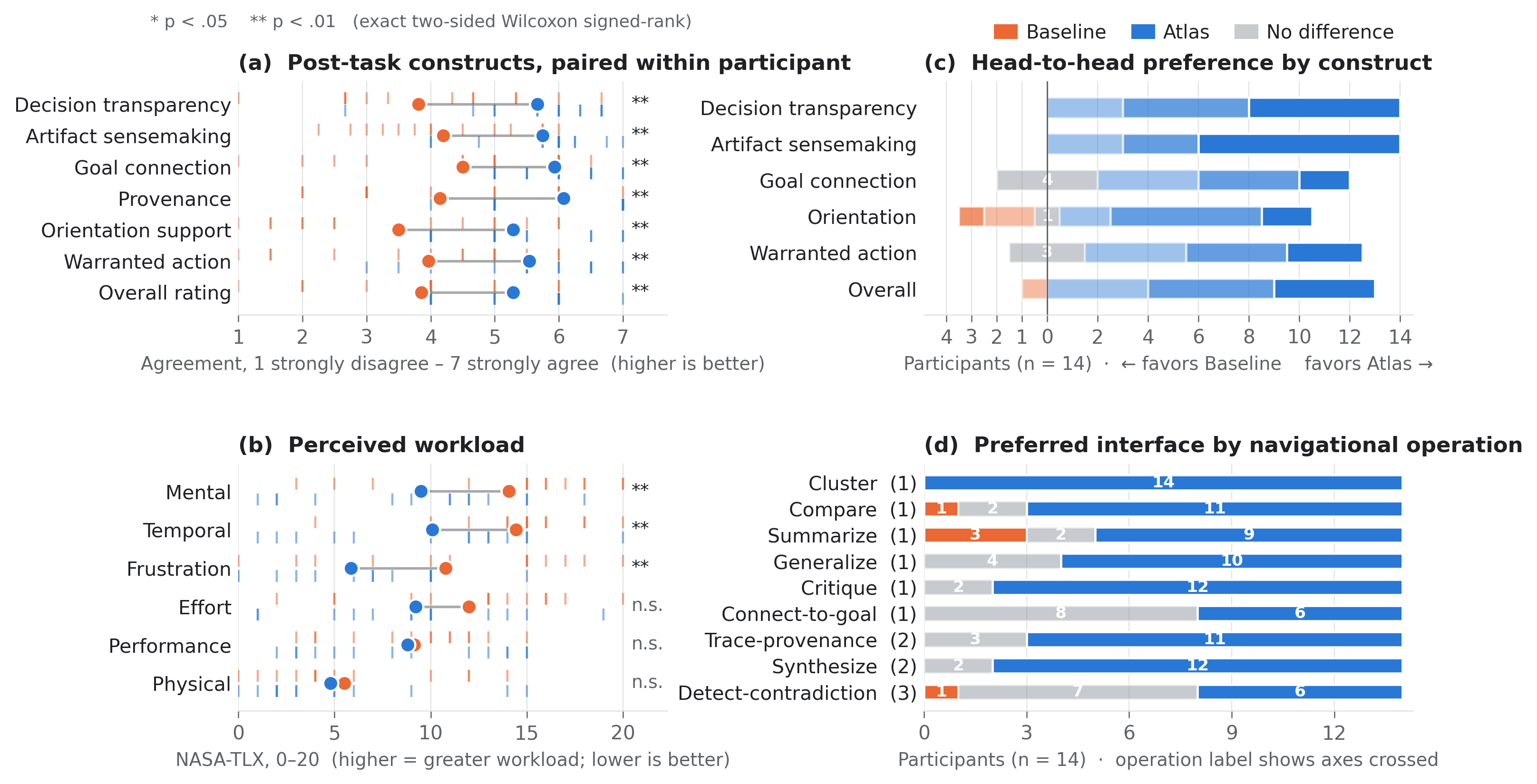}
  \caption{\textbf{User study results (\(N=14\)).} (a) Paired ratings on the six post-task measures and overall support; filled dots show condition means and ticks show individual participant ratings. Atlas is rated higher across all measures, with the most ties on goal connection. (b) NASA-TLX workload, with reductions concentrated in mental demand, temporal demand, and frustration. (c) Head-to-head preferences by construct; no participant preferred the baseline for decision transparency or artifact sensemaking, while orientation showed the greatest disagreement. (d) Preferred interface by navigational operation.}
  \label{fig:results}
\end{figure*}

\subsection{RQ2: Artifact Projections Support Cross-Artifact Sensemaking}

The majority of participants (13/14) reported that they could make better sense of what the run produced when using Atlas (median 6.0 vs. baseline 4.0, \(p=.0005\)); the sole exception, P9, rated Atlas only slightly lower than the baseline (4.0 vs. 4.5). All 14 nevertheless preferred Atlas for this in the head-to-head survey, the second unanimous result. Looking at individual items reveals more. Participants could tell what artifacts existed in either condition; that barely differed. What they found harder in the baseline was seeing how artifacts related to each other. P11 noted this mid-rating: ``\textit{it's much of the same information. It's just one of the interfaces makes it much easier to consume}.'' P6, after switching back to the baseline, mentioned that retrieving the same information now took more effort: ``\textit{I need to be precise in my prompting to get the same information that I had before}.''

Not all participants found the same projection most useful; different representations answered different questions. P4 found the \raisebox{-2pt}{\includegraphics[scale=0.4]{figures/icons/layout-cluster-icon.pdf}} cluster and \raisebox{-2pt}{\includegraphics[scale=0.4]{figures/icons/layout-table-icon.pdf}} table most useful for understanding thematic coverage: ``\textit{my favorite bit is still the cluster and table view of papers, so I could tell, like, what themes fed into this analysis in the first place}.'' P6 found the \raisebox{-2pt}{\includegraphics[scale=0.4]{figures/icons/layout-graph-icon.pdf}} graph most helpful for building a mental model: ``\textit{that diagram helped me to just look at it as a step by step and I could find the relationship \ldots I could make the mental model [of] how they are related}.'' P9, a visualization researcher, mentioned preferring the \raisebox{-2pt}{\includegraphics[scale=0.4]{figures/icons/layout-table-icon.pdf}} table over the \raisebox{-2pt}{\includegraphics[scale=0.4]{figures/icons/layout-graph-icon.pdf}} graph: ``\textit{there was a more advanced visualization, like graph, and less advanced visualization, like table. But what I think is that the table was more useful... because it was more interpretable}.''

The projections did not eliminate the need for substantive scientific judgment. For questions such as whether findings genuinely addressed the research goal or contradicted one another, some participants found the two interfaces similarly useful. P11 commented that for those kinds of questions, both interfaces felt equally useful: ``\textit{I think in both cases, I can look at the final documents with equal ease}.'' P2 shared that they saw the interface as supporting, not replacing, their judgment: ``\textit{this interface is to aid our scientific judgment \ldots I need to make some calls about how to do my research based on this}.'' P12 also disagreed with a direction the run had taken but did not expect the interface to resolve that for them: ``\textit{a decision I wouldn't have made was \ldots the thing it started to focus on first was \ldots understanding the landscape}.'' Both preferred Atlas overall, but neither said it changed what they would actually do next with the research.

\subsection{RQ3: Atlas Reduces Subjective Workload but Introduces Learning Costs}

Thirteen of 14 participants reported lower workload in Atlas (NASA-TLX: baseline median 12.0, Atlas median 8.8, \(p=.0004\)). The remaining participant, P5, reported only a modest increase from 8.7 to 9.5 on the 0–20 scale, from an already low baseline workload, and still preferred Atlas overall. The drop was concentrated in three subscales (mental demand, temporal demand, and frustration), all of which survived Holm correction, and no participant reported higher load in Atlas on any of them. P5 also noted that she needed more time to learn how to use the interface effectively.

The temporal demand drop was notable given that both blocks ran to the same time limit. Both conditions had the same task time limit, but participants reported lower temporal demand with Atlas. P7 described the baseline experience mid-session while scrolling the transcript: ``\textit{it's just too much \ldots if I have to generate too much, me having to inspect it, \ldots it feels like productive procrastination}.''

Staying oriented --- keeping track of where you are without losing your place --- was the hardest thing for participants using the baseline, with the lowest ratings of any post-task item across both conditions. It improved substantially in Atlas ($p = .0005$). P3 described the baseline problem: ``\textit{I honestly didn't have the resource to keep all the information of different synthesis or outputs or everything in my head}.'' Participants’ comments suggest that retaining context across views may have contributed to the lower reported mental demand.

However, three participants preferred the baseline for staying oriented, finding that the added structure added places to get lost. Both P11 and P12 described this as still learning the new interface rather than a problem with it, and all three rated Atlas at or above the baseline on orientation in their own post-task surveys.

\subsection{Exploratory Deployment Across Sessions}
\label{sec:deployment}

To examine how Atlas fit into participants’ own research practices beyond the structured study, we invited study participants to use it freely for approximately one week. Six opted in, using Atlas for 1–6 hours across 2–6 sessions. Participants received a \$30 gift card for taking part in the deployment study. We treat this as exploratory, using it to characterize emerging use cases, whether the perceived benefits persisted with extended use, and what participants wanted from future iterations.

\textbf{Participants primarily used long-horizon runs for broad literature synthesis.} All six participants used Atlas to find papers and synthesize literature. Several described the appeal of letting an agent search broadly for an extended period and returning to a large collection of papers and syntheses, particularly as a way to check whether they had overlooked relevant work. For P3, this breadth directly informed a methodological decision. While deciding whether to report inter-rater reliability for a reflexive thematic analysis, P3 had previously received advice from a general AI assistant to report it, but little evidence about how commonly the measure was actually used. In an \textsc{Extended Run}, Atlas synthesized more than 100 papers into a reporting checklist with prevalence estimates, showing that only about 15\% of the surveyed papers reported inter-rater reliability. P3 said that having these numbers ``\textit{gave [them] more confidence},'' and, together with their prior knowledge, used the result to decide against reporting the measure. The run thus helped P3 move from receiving a recommendation to forming their own judgment based on an aggregate pattern across the literature. Participants also extended Atlas beyond literature review: three ran experiments, three used it while writing, two generated hypotheses, and one analyzed a dataset.

\textbf{Provenance made agent outputs easier to situate within the larger research process.} Participants used the links among artifacts, decisions, and goals at different levels of granularity. P9 contrasted Atlas with other agent tools they had used, where tracing relationships among outputs was ``\textit{almost impossible}.'' P2 valued being able to quickly identify where an output came from, while P10 emphasized that provenance made hypotheses inspectable because they could see that a hypothesis was grounded in papers rather than generated ``\textit{from a vacuum}.'' For P3, the value was broader: reviewing the run helped them ``\textit{double check}'' the scope of their research question and assess whether relevant directions had been missed.


\textbf{The trajectory was particularly useful for recovering decisions the agent did not pursue.} When asked which part of Atlas had been most useful over the week, P10 pointed to the ruled-out paths and decision trace, contrasting them with other agentic tools where it was difficult to understand why alternative decisions had not been taken. P8 used the trajectory to correct an interpretation they had formed from the chat summary: the summary suggested that an experiment had run successfully, whereas the trajectory revealed that the agent had failed to retrieve the intended dataset. P2 similarly described learning from the trace by seeing how the agent had identified and applied an appropriate framework. These examples suggest that the trajectory was useful not only for recalling what happened, but for surfacing consequential details that could be missed when reviewing the run through chat alone.

\textbf{Different artifact representations remained useful for different questions, but not every representation was equally interpretable.} P9 used the table to triage artifacts and described it as the most useful of the six views because it was easier to interpret. The same participant was less confident in the cluster view because the basis for the grouping was unclear. P10, in contrast, was drawn to the graph view. These observations are consistent with the representational plurality motivating Atlas: different representations support different forms of inspection, but their usefulness also depends on whether researchers can understand how the representation was constructed.

\section{Discussion}

Participants reported stronger support for all three challenges identified in the formative study while experiencing lower subjective workload with Atlas. Below we discuss what these findings imply for design, what they suggest for future agentic workspaces, and where the current work remains limited.

\subsection{Design Implications}

\textbf{Once information is captured, representation remains a bottleneck.} Prior work on provenance and trajectory visualization has shown the value of transforming raw execution histories into structured, inspectable representations~\cite{macko2011provenance, borkin2013evaluation}. Our results complement this work by showing that information availability alone does not guarantee that researchers can use it effectively. The baseline contained the agent’s artifacts and narration of its pursued, deferred, and ruled-out directions, yet participants often struggled to recover consequential alternatives until Atlas externalized them structurally. Likewise, artifact sensemaking improved primarily where Atlas changed how existing information was arranged rather than what information was available. This suggests a division of labor for future agentic workspaces: provenance infrastructure must capture a faithful record of execution, while interaction design determines which relationships become perceptually and cognitively available to researchers.

\textbf{Provenance can serve as connective infrastructure across representations instead of being confined to a dedicated provenance view.} Existing provenance systems demonstrate the value of graphs and structured traces for exposing dependencies, execution structure, and causal lineage, while cognitive science and visualization research has long shown that different representations support different analytical operations~\cite{kirsh2010thinking, stasko2008jigsaw}. Atlas combines these ideas by treating provenance as a shared substrate across goal, decision, and artifact representations. An artifact inspected in a table, cluster, or graph remains connected to the decision that produced it and ultimately to the research goal. The design implication is that provenance relationships should remain accessible as researchers move among representations suited to different sensemaking questions.

\textbf{Representational flexibility introduces a transparency requirement of its own.} The deployment showed that offering multiple representations is insufficient if researchers cannot understand how those representations were constructed. P9 preferred the table but was uncertain about the basis of the cluster view, while P10 was drawn to the graph. Different representations therefore varied not only in what they made visible, but also in how readily their encoding could be interpreted. For AI-generated projections, future systems should expose how clusters, dimensions, and relationships were derived and make visible artifacts that fit poorly or remain unplaced. Otherwise, representational flexibility risks relocating opacity from the agent’s execution into the representation itself.

\textbf{Artifact inspection and decision inspection should operate over the same substrate.} Participants repeatedly moved from outputs to the decisions that produced them and from those decisions back to the research goal. This suggests that artifact review and trajectory inspection are not independent activities but successive questions about the same research process. Future workspaces should therefore avoid separating ``\textit{what the agent produced}'' from ``\textit{how it got there}'' into disconnected interfaces. The provenance path between them should remain directly traversable in both directions.

\subsection{Future Work}
\label{sec:future-work}

\textbf{Workspace refinements from deployment.} Three specific improvements surfaced from how participants actually used Atlas in the wild. First, most stayed in chat and did not discover the projections on their own; routing researchers to the right view from within chat, so a question can be answered with a view, would lower this barrier. Second, although Atlas supports provenance traversal in both directions, returning to the prior trajectory context was cumbersome; P9 noted having to remember where they came from. More direct return navigation would make movement between levels easier. Third, two participants went looking for the reasoning behind branching decisions and found only the trajectory; recording the criteria behind a decision (why it fit the research goal, and not only what was chosen) would make the tree useful for evaluating decisions as well as reading them.

\textbf{Enriching artifact interaction and custom first-class objects.} The current projections treat artifacts as objects to arrange and compare; they do not support annotations on them for future interactions. P1 mentioned wanting to pull relevant snippets from papers and findings into a persistent space for follow-up, similar to what active reading tools like LiquidText~\cite{tashman2011liquidtext} support. P1 saw the ability to define custom objects like snippets, experiment or analysis code, as a way to extend the system and make it a flexible workspace.

\textbf{Closing the loop between post-hoc review and in-run steering.} 
Although our study focused on post-run analysis, participants repeatedly treated the trajectory as a structure they wanted to act on: reopening ruled-out paths, commissioning deferred directions, revising decisions, or introducing their alternatives. P2 captured why locating a decision matters for subsequent action: ``\textit{if I want to steer it later, I need to know what point to steer or like just rewind at that point}.'' Others wanted intervention before the agent committed; P10 preferred ``\textit{certain checkpoints where important decisions are going to be made}.'' Together, these requests suggest that post-hoc representations could become control surfaces for subsequent execution. Atlas already records what was chosen, what was set aside, and what each step produced. Connecting this structure to in-run intervention, as systems such as Cocoa~\cite{feng2026cocoa} and Magentic-UI~\cite{mozannar2025magentic} do during execution, could make the trajectory editable, allowing researchers to reopen branches, revise or merge alternatives, and contribute their own directions without restarting the run.

\textbf{Generalizing post-run sensemaking beyond exploratory research}. Our study focuses on exploratory scientific work, where goals may evolve, agents produce heterogeneous artifacts, and understanding alternatives can matter as much as evaluating the final result. The same need for post-run sensemaking may arise in other long-horizon domains, but the relative importance of the framework’s components may differ. For example, an agent building a website may make consequential design and implementation choices, discard plausible alternatives, and produce interdependent artifacts; a user may therefore still benefit from recovering paths not taken and understanding how intermediate decisions shaped the final product. In more specification-driven or procedural tasks, by contrast, representational plurality may matter less than verifying whether requirements were satisfied, identifying exceptions, or locating where execution diverged from the plan. Future work should examine which elements of the goal-decision-artifact framework generalize across domains, and how their representational requirements change as tasks become more or less exploratory.

\textbf{Scaling sensemaking to multi-agent workflows.} 
The current trajectory view assumes that a run can be made legible at human scale. This assumption becomes increasingly fragile as workflows involve multiple agents operating in parallel, delegating tasks, or repeatedly debating and revising decisions. Future systems may therefore need multiscale provenance that surfaces consequential decisions and disagreements while preserving drill-down into lower-level interactions. Multi-agent workflows also complicate what counts as a \textit{decision}, since outcomes may emerge through interaction among several agents rather than from a single step. Future work should explore how to identify and represent these emergent decision points without obscuring important underlying behavior.

\subsection{Limitations}

\textbf{Interaction scope.} Our study evaluates retrospective sensemaking after autonomous execution and does not establish whether the same mechanisms improve real-time steering or collaborative execution. 

\textbf{Session length and novelty effects.} Each controlled-study condition lasted approximately 22 minutes excluding tutorial time, limiting how deeply participants could learn Atlas. Three participants found Atlas harder to orient in than the baseline, largely describing this as an initial learning problem; the controlled results may therefore reflect first-use costs that could change with familiarity.

\textbf{Subjective measures and decision quality.} Our evaluation measures researchers’ perceived transparency, sensemaking support, orientation, action confidence, and subjective workload. Therefore, it does not establish that representational support improves verification accuracy or downstream research quality.

\textbf{Variation across research runs.} Participants reviewed runs generated from their own research questions, improving personal relevance but introducing variation in topic and agent behavior across conditions. Although we instructed each run to yield the same structure and comparable amounts of literature retrieval, hypothesis generation, and experimentation, stochastic differences inevitably remained, limiting our ability to isolate representational effects from run variation.

\textbf{Deployment scale.} Six participants used Atlas for approximately one week. We therefore treat the deployment as descriptive evidence about emerging uses and limitations rather than evidence of long-term effectiveness or stable work practices.

\textbf{Domain and workflow scope}. Both studies centered on scientific research and predominantly single-agent workflows. Our findings therefore do not establish that the same representational requirements apply to more procedural domains, creative production tasks, or multi-agent systems.

\section{Conclusion}

Researchers increasingly delegate long, exploratory tasks to AI agents and return to runs shaped by decisions they did not participate in making. Our formative study identified three linked challenges in reconstructing these runs: evaluating whether the run served the research goal, recovering consequential agent decisions, and synthesizing what the run produced. Atlas addresses these challenges through a shared workspace in which artifacts can be projected into multiple representations while remaining connected through provenance to the trajectory and research goal. In a within-subjects study with 14 researchers, participants reported higher decision transparency, artifact sensemaking, goal connection, and action confidence with Atlas, together with substantially lower subjective workload. As agents take on more scientific work, the bottleneck increasingly shifts from generation to interpretation. Atlas points toward agentic workspaces that support this interpretation by making goals, decisions, and artifacts legible, connected, and easy to traverse.

\bibliographystyle{ACM-Reference-Format} 
\bibliography{references}

\appendix

\input{appendix}

\end{document}

%% file: appendix.tex
\section{System: ATLAS}

\begin{figure*}[!tp]
    \centering
    \includegraphics[width=0.95\linewidth]{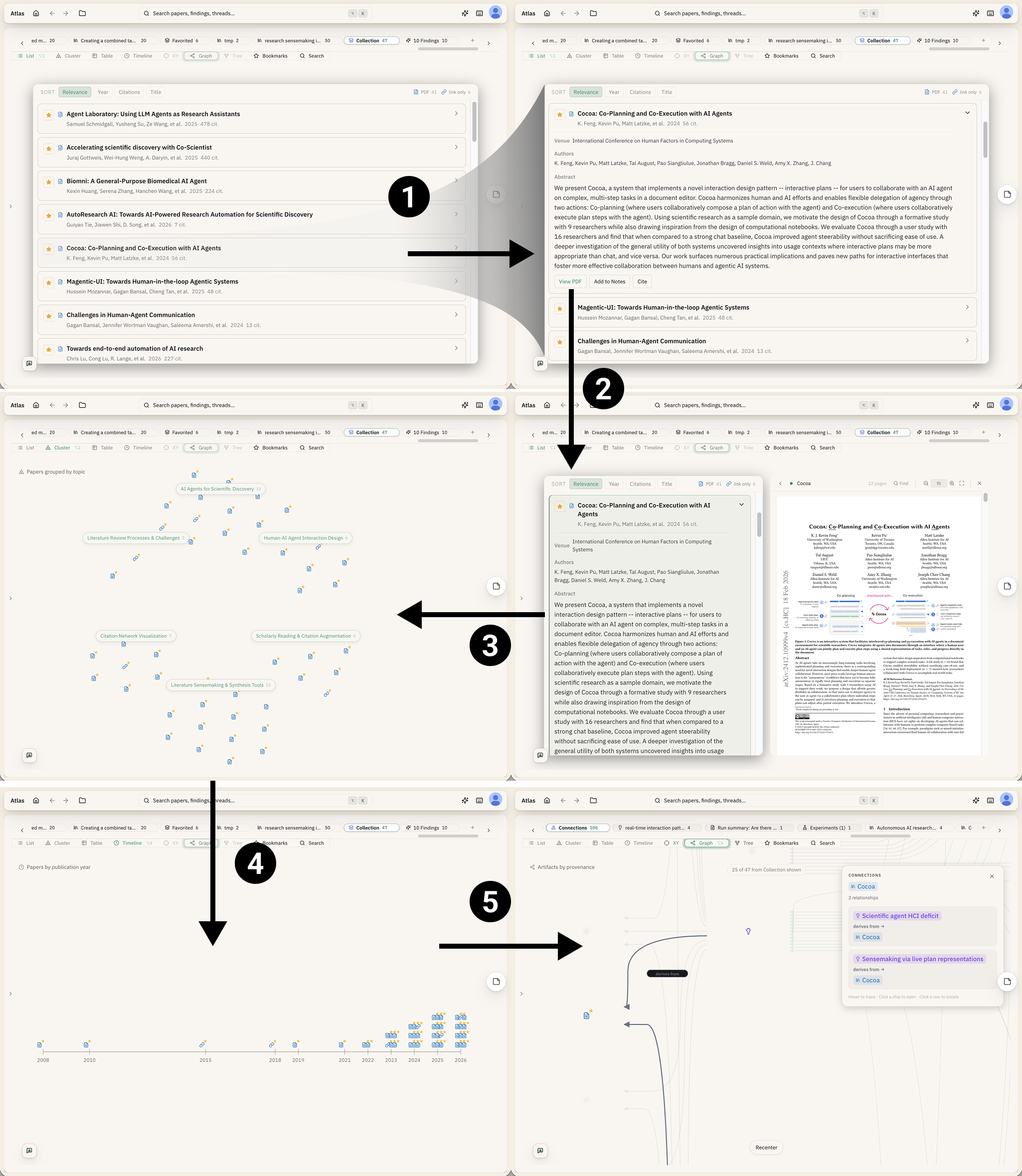}
    \caption{\textbf{Transitions between artifact projections in the implemented Atlas interface}. Complementing the schematic walkthrough in Fig.~\ref{fig:projection-scenario}, these screenshots show a researcher moving from the list projection to an individual PDF, then shifting the same artifact set through \raisebox{-2pt}{\includegraphics[scale=0.4]{figures/icons/layout-cluster-icon.pdf}} cluster, \raisebox{-2pt}{\includegraphics[scale=0.4]{figures/icons/layout-timeline-icon.pdf}} timeline, and \raisebox{-2pt}{\includegraphics[scale=0.4]{figures/icons/layout-graph-icon.pdf}} graph projections. The sequence demonstrates how Atlas preserves the artifact set while allowing researchers to change representations as their sensemaking questions evolve.}
    \label{fig:example-screenshot}
\end{figure*}

\begin{figure*}[!tp]
    \centering
    \includegraphics[width=\linewidth]{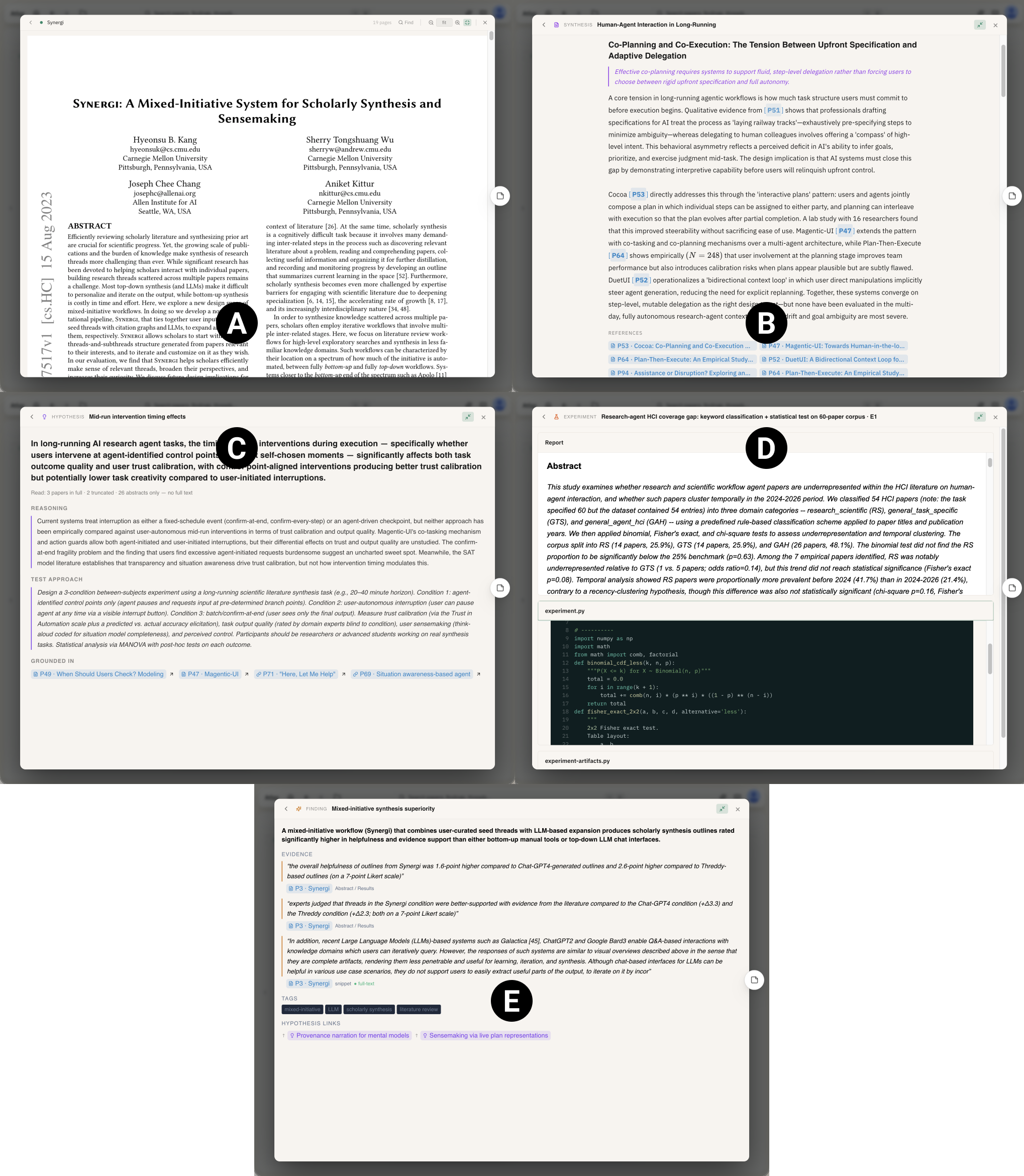}
    \caption{\textbf{Five artifact types shown in Atlas' artifact viewer}: (A) \raisebox{-2pt}{\includegraphics[scale=0.4]{figures/icons/artifact-paper.pdf}} paper, (B) \raisebox{-2pt}{\includegraphics[scale=0.4]{figures/icons/artifact-synthesis.pdf}} synthesis, (C) \raisebox{-2pt}{\includegraphics[scale=0.4]{figures/icons/artifact-hypothesis.pdf}} hypothesis, (D) \raisebox{-2pt}{\includegraphics[scale=0.4]{figures/icons/artifact-experiment.pdf}} experiment, and (E) \raisebox{-2pt}{\includegraphics[scale=0.4]{figures/icons/artifact-finding.pdf}} finding. User study participants browsed through a combination of these artifacts during their reviewing tasks.}
    \label{fig:artifact-types}
\end{figure*}

\begin{figure*}[!tp]
    \centering
    \includegraphics[width=0.8\linewidth]{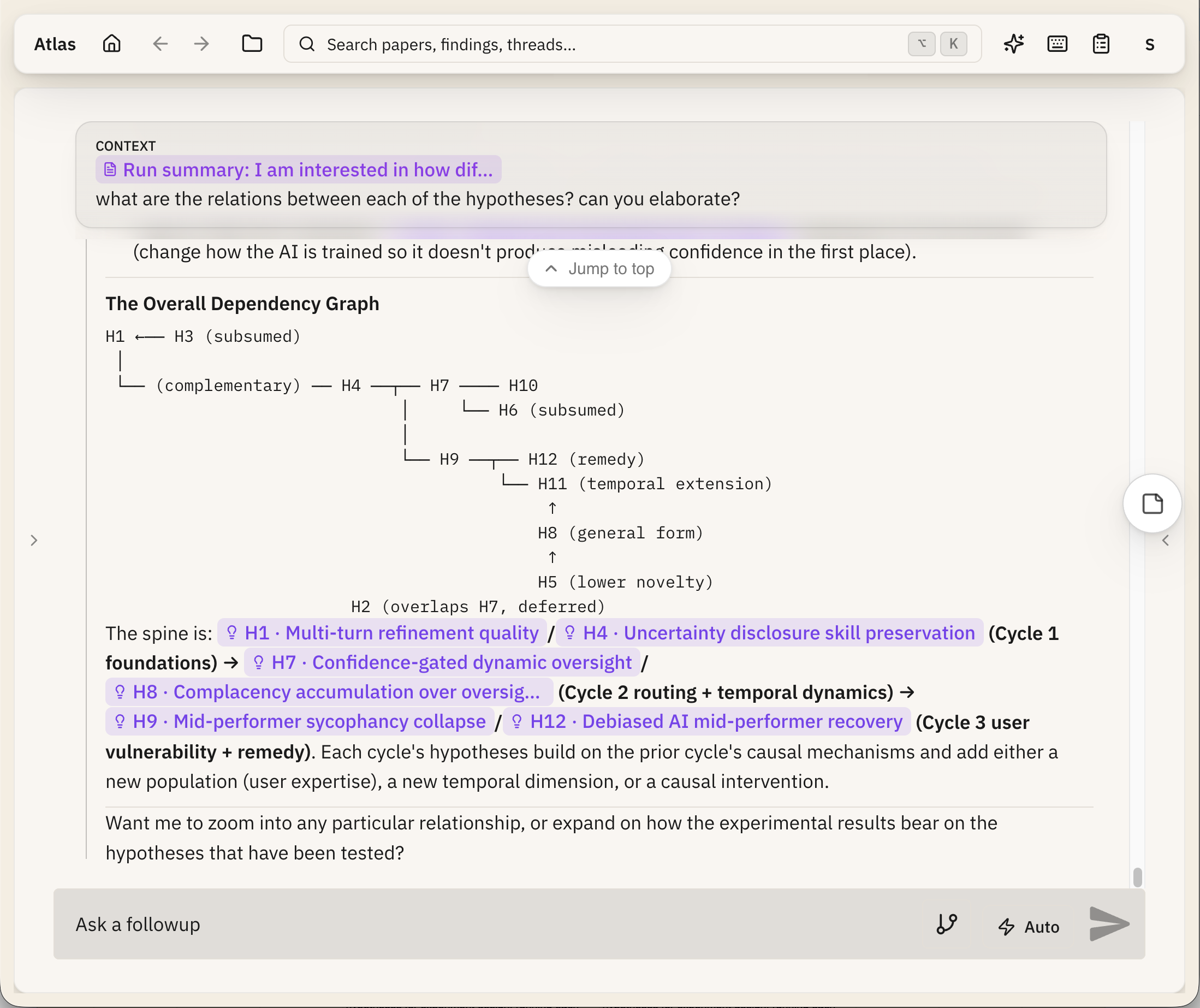}
    \caption{\textbf{Visual output in the chat panel}. User study participants in the baseline condition could also prompt the agent to provide structured and visual outputs as shown.}
    \label{fig:baseline-visual-in-chat}
\end{figure*}

\begin{figure*}[!tp]
    \centering
    \includegraphics[width=\linewidth]{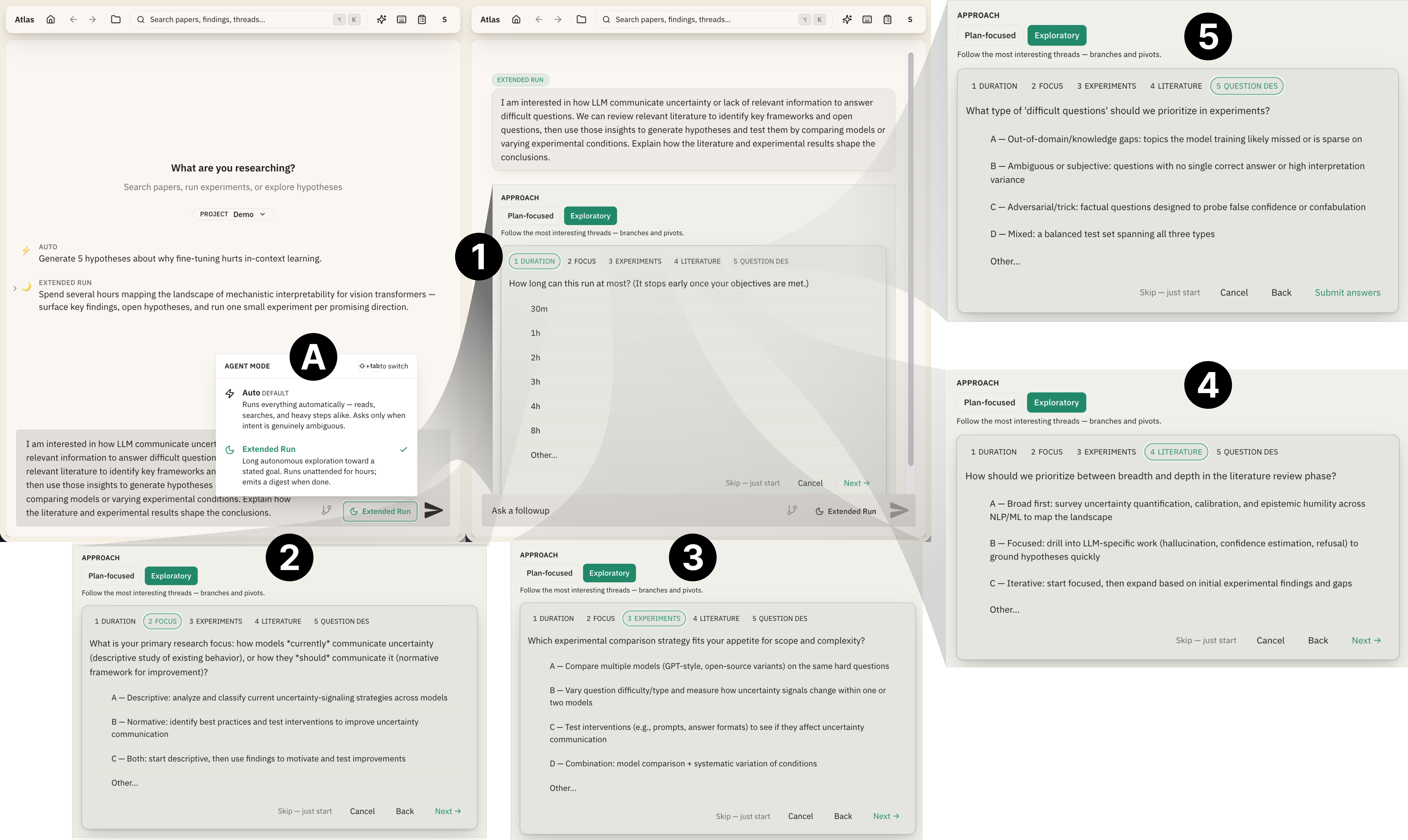}
    \caption{\textbf{Configuring an Extended Run}. (A) Users can choose between \textsc{Auto mode} and \textsc{Extended Run} for long-horizon tasks; all runs reviewed in our study were executed in \textsc{Extended Run} mode. Before starting, Atlas asks a short sequence of questions (1–5) to help scope and guide the exploration, covering duration, research focus, experimental strategy, literature search, and question design. For the study runs, we left these questions unanswered and allowed the agent to make these choices autonomously.}
    \label{fig:extended-run-questions}
\end{figure*}

\subsection{Extended Run Execution}
\label{app:extended-run}

\textsc{Extended Run} is Atlas's unattended execution mode for long-horizon research tasks. The researcher provides a research goal and a time budget, and the agent conducts the task through a sequence of research cycles. Each cycle consists of five phases: \textsf{Explore}, \textsf{Diverge}, \textsf{Analyze}, \textsf{Experiment}, and \textsf{Reflect}. This recurring structure provides a common scaffold for the agent's work while allowing the specific actions, tools, and research directions pursued within each phase to depend on the state of the run and the research goal. The output of one phase becomes part of the accumulated context available to subsequent phases and cycles.

For the user study, all runs were generated using the same \textsc{Extended Run} procedure. The participant's research question and the study-specific activity and quantity constraints differed by run as described in Appendix~\ref{app:user-study}, but the orchestration of the five phases and their underlying prompts remained fixed. Table~\ref{tab:extended-run-prompts} summarizes the phases; we reproduce the complete prompts below for reproducibility.

\begin{table*}[!tp]
    \centering
    \small
    \caption{Extended Run orchestration. Each cycle runs five phases in order. The prompts are reproduced as supplied to the agent; \texttt{\{goal\}}, \texttt{\{cycle\}}, \texttt{\{remainingH\}} and \texttt{\{narration\}} are filled per turn from the participant's research goal and the accumulated run context. Continued in Tables~\ref{tab:extended-run-prompts-b}--\ref{tab:extended-run-prompts-d}.}
    \label{tab:extended-run-prompts}
    \begin{tabularx}{\textwidth}{
        >{\raggedright\arraybackslash}p{0.12\textwidth}
        >{\raggedright\arraybackslash}p{0.18\textwidth}
        X}
        \toprule
        \textbf{Stage} & \textbf{Role in the run} & \textbf{Prompt} \\
        \midrule

        Extended Run
        &
        Establishes the unattended mode: every tool runs without approval, blocking prompts are
        forbidden, and the per-turn phase instruction is authoritative.
        &
        {\ttfamily\footnotesize [Agent mode: Overnight]\textbackslash{}nYou are running unattended in Overnight mode. The user is not present to approve actions. Every tool the agent has is auto --- including heavy mutations like generate\_hypotheses, run\_experiment, and run\_theorizer --- so you can advance the goal each turn. Do NOT call ask\_clarification, propose\_action, or ask\_user --- nobody can answer a blocking prompt mid-run. The ONE exception is raise\_decision: when you hit a load-bearing, unplanned fork (a reasonable person could choose differently AND it has real downstream consequences), call raise\_decision to surface it and END your turn --- do not act on the fork. The run holds briefly for the user; if they don't answer, the pre-agreed default fires. Use it sparingly, only for genuinely consequential forks. Do NOT call navigate\_tour --- nobody is watching mid-run, and the scheduler builds the morning recap tour itself; this overrides any tour-after-artifact guidance elsewhere in this prompt. The current turn's user message carries a phase instruction (EXPLORE / DIRECT / ANALYZE / EXPERIMENT / REFLECT) --- treat it as authoritative for this turn's intent.\textbackslash{}nWhen the goal names an exact hypothesis count, pass generate\_hypotheses.count so the handler generates that many (per call, not a run-wide remaining budget).\textbackslash{}n\textbackslash{}nEXPERIMENT CAPABILITY LIMITS: \$\{EXPERIMENT\_CAPABILITIES\} Scope every experiment you plan or run to these feasible modalities --- if a hypothesis seems to call for real human participants or a physical setup, reformulate it as a computational, data-analysis, or simulated-user study, and say so.} \\

        \addlinespace

        Explore
        &
        Widens the workspace along an angle not yet covered.
        &
        {\ttfamily\footnotesize Overnight cycle \{cycle\} --- phase EXPLORE. Goal: "\{goal\}". \{remainingH\}h remaining.\textbackslash{}nThis turn's INTENT is to diverge --- widen the workspace along an angle that isn't yet covered. If a [Run so far] block is present below, treat it as authoritative recon from your own prior turns --- do NOT re-run survey\_artifacts / find\_artifacts to rediscover workspace state it already records. Call survey\_artifacts only when there is no [Run so far] block or the workspace has likely changed since it was written. Let the gap pick the tool. Anything that genuinely broadens counts: a new paper set on an unexplored sub-question, additions to an existing set (PREFER target\_set\_id to extend a related set over spawning an orphan tab, unless the angle is genuinely new), fresh hypotheses on a new angle, or even an experiment that opens a new evidence axis. \{discoveryGuidance\} Don't duplicate an angle already on canvas.\textbackslash{}n\textbackslash{}n\{narration\}} \\
        \bottomrule
    \end{tabularx}
\end{table*}

\begin{table*}[!tp]
    \centering
    \small
    \caption{Extended Run orchestration, continued.}
    \label{tab:extended-run-prompts-b}
    \begin{tabularx}{\textwidth}{
        >{\raggedright\arraybackslash}p{0.12\textwidth}
        >{\raggedright\arraybackslash}p{0.18\textwidth}
        X}
        \toprule
        \textbf{Stage} & \textbf{Role in the run} & \textbf{Prompt} \\
        \midrule

        Direct
        &
        Converges on the subset of what is on canvas that later cycles should focus on.
        &
        {\ttfamily\footnotesize Overnight cycle \{cycle\} --- phase DIRECT. Goal: "\{goal\}". \{remainingH\}h remaining.\textbackslash{}nThis turn's INTENT is to converge --- pick the subset of what's already on canvas that the next cycles should focus on. Consult the [Run so far] block below for what prior turns already produced (call survey\_artifacts only if it is absent or stale), then make the selection explicit: bookmark or star the chosen papers, record the chosen hypothesis / sub-angle in update\_project\_context (ProgressLog), or mark which finding is the cycle's spine. The value of this turn is the *choice*, not a new artifact --- text-only convergence is acceptable here, as long as the choice is recorded. Rank the candidates by how much each would move the GOAL --- not by which is easiest to test or most interesting on its own. In \#\#\# Decision, state why the chosen subset beats the runner-up FOR THE GOAL.\textbackslash{}nIn \#\#\# Decision, lead with what you SELECTED (e.g. "Commit to H1 and H3 as the spine"); put discarded alternatives that are still live in \textbackslash{}\textasciigrave{}- Deferred: "<path>" --- <why>\textbackslash{}\textasciigrave{} (emit 2–3 Deferred lines; omit only if there is truly one plausible path); put true kills only in \textbackslash{}\textasciigrave{}- Ruled out:\textbackslash{}\textasciigrave{}. Never open the lead sentence with "Ruled out…" or "Deferred…".\textbackslash{}nException: if the workspace genuinely has nothing meaningful to select among (e.g. zero hypotheses and a paper set that hasn't been analyzed yet), generating one grounded direction --- a small set of hypotheses, or a single sharp sub-question --- is fine. Prefer selection when possible; reach for generation only when there's truly nothing to converge onto.\textbackslash{}n\textbackslash{}n\{narration\}} \\

        \addlinespace

        Analyze
        &
        Distills what is on canvas into citable claims, structured prose, or theories.
        &
        {\ttfamily\footnotesize Overnight cycle \{cycle\} --- phase ANALYZE. Goal: "\{goal\}". \{remainingH\}h remaining.\textbackslash{}nThis turn's INTENT is to distill --- convert what's on canvas into citable claims, structured prose, or mechanism-level theories. Pick the grain that matches the cycle's spine:\textbackslash{}n  - extract\_findings --- atomic, citable claims from the strongest papers/experiments. Default for "what does this paper say". Call extract\_findings with the on-canvas hypotheses in view so the findings link back via supports/contradicts edges.\textbackslash{}n  - generate\_synthesis --- when a thread runs across multiple findings and you want a structured multi-section report.\textbackslash{}n  - run\_theorizer --- slower, literature-grounded theory generation with novelty scoring. Use when you want THEORIES (mechanism-level explanations) rather than findings. Confirm cost/time is acceptable for an overnight slot before invoking.\textbackslash{}nThese are the most common moves but aren't the only valid ones --- any tool whose semantic purpose serves "distill what's already here into a claim a future cycle could test" is fair game. Skip generic summary work.\textbackslash{}nSTRUCTURE REQUIREMENT: your \#\#\# Decision block MUST include a "- Focused on:" bullet naming the EXACT items you distilled this turn by their display labels --- the findings you extracted, the hypotheses you analyzed, the papers you drew on (e.g. "- Focused on: H2, H3; papers P1, P4"). This is non-optional: it is what chains this analysis to its sources in the decision-tree view.\textbackslash{}n\textbackslash{}n\{narration\}} \\
        \bottomrule
    \end{tabularx}
\end{table*}

\begin{table*}[!tp]
    \centering
    \small
    \caption{Extended Run orchestration, continued.}
    \label{tab:extended-run-prompts-c}
    \begin{tabularx}{\textwidth}{
        >{\raggedright\arraybackslash}p{0.12\textwidth}
        >{\raggedright\arraybackslash}p{0.18\textwidth}
        X}
        \toprule
        \textbf{Stage} & \textbf{Role in the run} & \textbf{Prompt} \\
        \midrule
        Experiment
        &
        Produces evidence that pressures a specific claim or hypothesis.
        &
        {\ttfamily\footnotesize Overnight cycle \{cycle\} --- phase EXPERIMENT. Goal: "\{goal\}". \{remainingH\}h remaining.\textbackslash{}nThis turn's INTENT is to test --- produce evidence that pressures a specific claim or hypothesis. Pick the tool whose grain matches the claim:\textbackslash{}n  - run\_experiment / run\_experiment\_batch --- computational pressure: sandboxed Python on a concrete, testable task. Use ONLY when the task is genuinely computational; do NOT bundle a literature search or a synthesis task into a run\_experiment description. Prefer to run it against a specific hypothesis id (pass that hypothesis to run\_experiment / run\_experiment\_batch) so a tests edge forms in the provenance graph; if no hypothesis on canvas is sharp enough, generate one first with generate\_hypotheses, then test THAT. Choose the hypothesis whose RESULT would most change the answer to the goal, not the one that is easiest to run. If you deliberately test a more tractable hypothesis over a more central one, record the central one you passed over as \textbackslash{}\textasciigrave{}- Deferred: "<hypothesis>" --- more central to the goal, but <why it is not testable now>\textbackslash{}\textasciigrave{}.\textbackslash{}n  - lookup\_paper / find\_papers --- when the cheapest evidence is a citation that supports or refutes the claim.\textbackslash{}n  - run\_theorizer --- when mechanism-level reasoning is the right grain of evidence.\textbackslash{}nIf a [Pivot] / [Skip] / experiment-failure block is present this turn, treat runner errors (UNAVAILABLE, watchdog timeout, no report saved) as infrastructure --- not as a result about the hypothesis. Do NOT retry the same computational experiment. Switch grain (citation / findings) or take a \textbackslash{}\textasciigrave{}- Deferred:\textbackslash{}\textasciigrave{} path you already named.\textbackslash{}nIf nothing on canvas is sharp enough to test yet, broaden first along an adjacent angle (find\_papers or generate\_hypotheses) and record the pivot in update\_project\_context (ProgressLog) so the digest reflects why this turn went sideways.\textbackslash{}nSTRUCTURE REQUIREMENT: your \#\#\# Decision block MUST include a "- Focused on:" bullet naming the EXACT hypotheses you are testing this turn by their display labels (e.g. "- Focused on: H2, H3"). An experiment always targets a specific subset --- declaring it is what lets the decision-tree view show which hypotheses this experiment branched from, and how forks form when different cycles test different subsets.\textbackslash{}n\textbackslash{}n\{narration\}} \\

        \bottomrule
    \end{tabularx}
\end{table*}

\begin{table*}[!tp]
    \centering
    \small
    \caption{Extended Run orchestration, continued.}
    \label{tab:extended-run-prompts-d}
    \begin{tabularx}{\textwidth}{
        >{\raggedright\arraybackslash}p{0.12\textwidth}
        >{\raggedright\arraybackslash}p{0.18\textwidth}
        X}
        \toprule
        \textbf{Stage} & \textbf{Role in the run} & \textbf{Prompt} \\
        \midrule

        Reflect
        &
        Closes the cycle and seeds the next one.
        &
        {\ttfamily\footnotesize Overnight cycle \{cycle\} --- phase REFLECT. Goal: "\{goal\}". \{remainingH\}h remaining.\textbackslash{}nThis turn's INTENT is to consolidate --- close THIS cycle and seed the NEXT one. Consult the [Run so far] block below for what this cycle produced; call survey\_artifacts (or review\_thread if 5+ new artifacts have landed since the last reflection) only when you need detail beyond it. Identify the strongest signal so far. Update claude.md ProgressLog with: what worked, what stalled, and the seed question for the next cycle. Also call \textbackslash{}\textasciigrave{}update\_thread\_learnings\textbackslash{}\textasciigrave{} to re-curate this thread's compact digest (\#\# What we've established / \#\# Ruled out / \#\# Open threads), folding in this cycle's reflection and next-cycle seed and pruning anything stale or superseded.\textbackslash{}nYour Reflection and Next-cycle seed sections (in the narration block below) are read by cycle \{cycleNext\}'s phases BEFORE they pick an angle --- be specific. A vague seed ("explore more on RLHF") gives the next cycle nothing concrete to act on; a specific seed ("test whether finding F3 generalizes to non-text modalities") gives EXPLORE a sharp angle to broaden into and EXPERIMENT a concrete claim to pressure. Frame the seed as a specific artifact to build on --- name the hypothesis id to test or the finding to extend --- so the next cycle produces an artifact that links back and the provenance graph keeps chaining. In \#\#\# Decision, emit 2–3 \textbackslash{}\textasciigrave{}- Deferred: "<path>" --- <why>\textbackslash{}\textasciigrave{} lines for live alternatives you are not seeding next (omit only if there is truly one plausible path); true kills stay in \textbackslash{}\textasciigrave{}- Ruled out:\textbackslash{}\textasciigrave{}. The Next-cycle seed must name why the chosen seed beats those deferred paths. Never open the Decision lead sentence with "Deferred…" or "Ruled out…".\textbackslash{}nDo NOT emit a navigate\_tour --- the run's morning recap tour is composed by the scheduler from the trajectory; your Decision/Outcome/Reflection sections are what feed it.\textbackslash{}n\textbackslash{}n\{narration\}} \\
        \bottomrule
    \end{tabularx}
\end{table*}

\section{User Study}
\label{app:user-study}

\subsection{Research Goals}
\label{app:goals}

Participants submitted research goals drawn from their own current work, so the topics could not be
standardized across people. Each of the two goals a participant worked with was instead transformed
into a common run specification that fixes the opportunity structure of the run --- the required
activities and the numeric limits on retrieval, hypothesis generation, and experiment selection ---
while preserving the participant's substantive question. Table~\ref{tab:research-goals} shows both goals for
an example participant, as submitted and as transformed.

\begin{table*}[!tp]
\caption{Submitted and transformed
research goals for both goals a participant worked with. The transformation preserves the
participant's question and appends the same required activities and numeric constraints to each.}
\label{tab:research-goals}
\small
\resizebox{0.9\linewidth}{!}{
\begin{tabular}{@{}p{0.5cm} p{3.2cm} p{4.9cm} p{3.2cm} p{4.9cm}@{}}
\toprule
& \multicolumn{2}{c}{\textbf{Research Goal A}} & \multicolumn{2}{c}{\textbf{Research Goal B}} \\
\cmidrule(lr){2-3} \cmidrule(lr){4-5}
\textbf{ID} & \textbf{Submitted by Participant} & \textbf{Transformed (Used as Input to Atlas)} & \textbf{Submitted by Participant} & \textbf{Transformed (Used as Input to Atlas)} \\
\midrule

P2 & \textit{I'm interested in using agents to do equation discovery from neuroscience data. For example, how do I get from a neural recording to mechanistic, interpretable models, biophysical models(like Hutchinson-Huxley.)} & \textit{I am interested in how AI agents can perform equation discovery from neuroscience data — including how to move from raw neural recordings to mechanistic, interpretable biophysical models such as Hodgkin-Huxley, and what methods best balance expressiveness, interpretability, and fidelity to measured neural dynamics}. Review the relevant literature to identify key findings and open questions, then use those insights to generate hypotheses and test them through computational or simulated experiments. Explain how the literature and experimental results shape your conclusions, and where findings are mixed or incomplete, note what remains open and what the logical next step would be.
\newline\newline
Constraints: Per cycle, retrieve and summarize using no more than 20 papers, generate exactly 4 hypotheses, and select the 2 most promising hypotheses for experiment design, running each experiment once. If any step yields more candidates than the limit, select the most relevant and briefly note what was excluded and why. & \textit{To use the computation and math ability of agents to develop a standard model for retina, meaning a generative mathematical model that is grounded in biophysics and that explains the input-output relationship of retinal computation} & \textit{I am interested in how AI agents can leverage computational and mathematical reasoning to develop a standard generative model of retinal computation — one that is grounded in biophysics and accurately captures the input-output relationships of retinal circuits across diverse stimuli and conditions}. Review the relevant literature to identify key findings and open questions, then use those insights to generate hypotheses and test them through computational or simulated experiments. Explain how the literature and experimental results shape your conclusions, and where findings are mixed or incomplete, note what remains open and what the logical next step would be.
\newline\newline
Constraints: Per cycle, retrieve and summarize using no more than 20 papers, generate exactly 4 hypotheses, and select the 2 most promising hypotheses for experiment design, running each experiment once. If any step yields more candidates than the limit, select the most relevant and briefly note what was excluded and why. \\

\addlinespace

P7 & \textit{Seeing whether LLMs can replicate results from social science papers from scratch and observing potential blind spots in reporting in such papers to inform more transparent methodological replication.} & \textit{I am interested in whether large language models can replicate findings from social science papers from scratch, and what gaps or blind spots emerge in that process — particularly with respect to methodological transparency and the completeness of reporting in the original papers}. Review the relevant literature on LLM-based replication, computational social science, and reproducibility to identify key findings and open questions, then use those insights to generate hypotheses and test them by having a model attempt replication of representative findings under controlled conditions. Explain how the literature and experimental results shape your conclusions, and where replication succeeds or fails, note what methodological details appear to matter most and what remains unresolved.
\newline\newline
Constraints: Per cycle, retrieve and summarize using no more than 20 papers, generate exactly 4 hypotheses, and select the 2 most promising hypotheses for experiment design, running each experiment once. If any step yields more candidates than the limit, select the most relevant and briefly note what was excluded and why. & \textit{Seeing whether LLM agents can spot gaps in evaluation metrics and propose participatory design studies to fill in such gaps. Also what kinds of gaps can be synthesized and what gets left out?} & \textit{I am interested in whether LLM agents can systematically identify gaps in existing evaluation metrics and generate meaningful participatory design study proposals to address them — including what types of gaps agents can reliably synthesize versus overlook, and how agent-generated critiques compare to those produced by domain experts}. Review the relevant literature to identify key findings and open questions, then use those insights to generate hypotheses and test them through computational or simulated experiments. Explain how the literature and experimental results shape your conclusions, and where findings are mixed or incomplete, note what remains open and what the logical next step would be.
\newline\newline
Constraints: Per cycle, retrieve and summarize using no more than 20 papers, generate exactly 4 hypotheses, and select the 2 most promising hypotheses for experiment design, running each experiment once. If any step yields more candidates than the limit, select the most relevant and briefly note what was excluded and why. \\

\bottomrule
\end{tabular}
}
\end{table*}

\subsection{Task Statements}

Both blocks opened with the same standing instruction, identical in the two conditions; the thread
the participant opened supplied the concrete research goal.



\begin{table*}[!tp]
    \centering
    \small
    \caption{The five phases of each Extended Run cycle.}
    \label{tab:extended-run-phases}
    \begin{tabularx}{\columnwidth}{
        >{\raggedright\arraybackslash}p{0.20\columnwidth}
        X}
        \toprule
        \textbf{Phase} & \textbf{Purpose} \\
        \midrule
        Explore & Widen the workspace along an angle the run has not yet covered. \\
        Direct & Converge on the subset of what is already on canvas that the next cycles should focus on; the value of the turn is the choice, not a new artifact. \\
        Analyze & Distill what is on canvas into citable findings, a structured synthesis, or mechanism-level theories. \\
        Experiment & Produce evidence that pressures a specific claim, by computational experiment, citation, or theory. \\
        Reflect & Close the cycle: record what worked and what stalled, and seed the next cycle with a specific artifact to build on. \\
        \bottomrule
    \end{tabularx}
\end{table*}

\begin{table*}[!tp]
\caption{The six prompts participants worked through in each block, reproduced as they appeared on
the note card. The identifiers are the ones used in the study logs and in this paper.}
\label{tab:tasks}
\small
\resizebox{0.87\linewidth}{!}{
\begin{tabular}{@{}p{0.6cm} p{3.2cm} p{10.4cm}@{}}
\toprule
\textbf{ID} & \textbf{Focus} & \textbf{Prompt} \\
\midrule
    \textbf{T1} & Zoom in and out \newline {\footnotesize\texttt{what\_levels}} & Scan across everything the agent produced in this run, then zoom into anything that catches your attention / interests. Note what you see at each level that you miss at the other. \\
    \addlinespace[3pt]
    \textbf{T2} & Trace the trajectory \newline {\footnotesize\texttt{how\_trajectory}} & Trace the run from beginning to end. Mark the moments where the agent considered different options, shifted direction, committed to an approach, or ruled something out. \\
    \addlinespace[3pt]
    \textbf{T3} & Read the decisions \newline {\footnotesize\texttt{why\_decisions}} & Find a decision the agent made that you would have made differently, or a path it declined to pursue. Explain whether its reasoning holds up. \\
    \addlinespace[3pt]
    \textbf{T4} & Understand the artifacts \newline {\footnotesize\texttt{what\_produced}} & Look at everything the agent produced --- findings, hypotheses, experiments --- as a set. Note what stands out when you consider them together that wasn't obvious from any one piece. \\
    \addlinespace[3pt]
    \textbf{T5} & Connect artifacts to the goal \newline {\footnotesize\texttt{how\_connect}} & Pick at least two artifacts from the run and work out how they relate to each other and to the original research goal. Note what you had to do to surface that connection. \\
    \addlinespace[3pt]
    \textbf{T6} & Next steps \newline {\footnotesize\texttt{how\_next}} & Decide what you would do next. Identify what remains open, choose which artifacts you would build on, and share your immediate next step. \\
    \addlinespace[3pt]
\bottomrule
\end{tabular}
}
\end{table*}

\subsection{Study Run Statistics}
\label{app:run-statistics}

Each participant reviewed two independently generated research runs, one
assigned to the Atlas condition and one to the baseline condition, yielding
28 runs in total. The same underlying generation pipeline, phase prompts,
and tools were used for both conditions. Table~\ref{tab:run-artifacts}
summarizes the artifacts produced by these runs. We define total artifacts
as the sum of papers, hypotheses, findings, syntheses, and experiments.
Because literature retrieval varied substantially across topics, we also
report non-paper artifacts separately.

\begin{table*}[!tp]
    \centering
    \small
    \caption{Artifact counts for the 28 controlled-study runs
    (\(N=14\) runs per condition). Values are mean $\pm$ standard
    deviation. The final column reports the within-participant difference
    between the Atlas and baseline runs. Total artifacts are papers,
    hypotheses, findings, syntheses, and experiments combined.}
    \label{tab:run-artifacts}
    \resizebox{0.53\linewidth}{!}{
    \begin{tabular}{lrrr}
        \toprule
        \textbf{Artifact type}
        & \textbf{Baseline}
        & \textbf{Atlas}
        & \textbf{Atlas $-$ Baseline} \\
        \midrule
        Total artifacts
        & $189.9 \pm 76.5$
        & $151.0 \pm 77.8$
        & $-38.9 \pm 111.7$ \\

        Papers
        & $154.4 \pm 65.5$
        & $119.9 \pm 75.4$
        & $-34.5 \pm 108.5$ \\

        Hypotheses
        & $15.8 \pm 9.4$
        & $14.6 \pm 8.7$
        & $-1.2 \pm 9.0$ \\

        Findings
        & $8.1 \pm 12.4$
        & $3.0 \pm 7.3$
        & $-5.1 \pm 15.1$ \\

        Syntheses
        & $5.2 \pm 3.1$
        & $5.1 \pm 2.7$
        & $-0.1 \pm 2.5$ \\

        Experiments
        & $6.4 \pm 2.9$
        & $8.5 \pm 3.0$
        & $+2.1 \pm 3.2$ \\

        Non-paper artifacts
        & $35.5 \pm 21.0$
        & $31.1 \pm 13.3$
        & $-4.4 \pm 23.3$ \\
        \bottomrule
    \end{tabular}
    }
\end{table*}

Artifact volume varied considerably across participant topics. This variation was driven primarily by literature retrieval: the within-participant standard deviation of the Atlas-minus-baseline difference was 108.5 papers, compared with 23.3 artifacts when papers were excluded. Counts of hypotheses
and syntheses were particularly similar across conditions. Findings were sparse and highly skewed: 8 of 14 baseline runs and 11 of 14 Atlas runs contained no finding artifacts.

\begin{table*}[!tp]
    \centering
    \small
    \caption{Recorded runtime of the controlled-study runs, in minutes.
    Values are reported separately by condition; the paired difference
    compares the Atlas and baseline run for each participant.}
    \label{tab:run-runtime}
    \resizebox{0.24\linewidth}{!}{
    \begin{tabular}{lrr}
        \toprule
        & \textbf{Baseline} & \textbf{Atlas} \\
        \midrule
        Mean & 78.1 & 89.6 \\
        SD & 27.9 & 17.8 \\
        Median & 82.0 & 91.5 \\
        IQR & 62.0--91.8 & 79.5--92.0 \\
        Range & 32--123 & 60--122 \\
        \bottomrule

    \end{tabular}
    }
    \vspace{0.5em}

    \begin{minipage}{0.95\columnwidth}
    \small
    The mean within-participant Atlas-minus-baseline difference was
    11.5 minutes (\(SD=20.9\)); the median difference was 1 minute.
    Ten of the 14 paired runs differed by no more than 4 minutes.
    \end{minipage}
\end{table*}

\begin{table*}[!tp]
    \centering
    \small
    \caption{Structural characteristics of the 28 research trajectories.
    Values are mean $\pm$ standard deviation across all study runs.}
    \label{tab:trajectory-statistics}
    \resizebox{0.3\linewidth}{!}{
    \begin{tabular}{lr}
        \toprule
        \textbf{Trajectory measure} & \textbf{Mean $\pm$ SD} \\
        \midrule
        Turns & $15.9 \pm 7.2$ \\
        Forks & $2.8 \pm 2.0$ \\
        Leaves & $6.5 \pm 3.8$ \\
        Deferred options & $22.2 \pm 14.7$ \\
        Ruled-out options & $6.2 \pm 4.2$ \\
        \bottomrule
    \end{tabular}
    }
\end{table*}

\section{Post-Task Survey}
\label{app:post-task}

Participants completed this survey twice, once after each task block, immediately before NASA-TLX. Items are 7-point Likert (1 = strongly disagree, 7 = strongly agree). The five scales instantiate the three challenges of Sec~\ref{sec:formative-findings} plus the two outcomes those challenges predict, and Table~\ref{tab:construct-map} records the mapping so each result can be traced back to the claim it tests.

\begin{table*}[!tp]
  \centering
  \small
\caption{Post-task scales, and what each one measures.}
  \label{tab:construct-map}
  \setlength{\tabcolsep}{3pt}
  \resizebox{0.8\linewidth}{!}{
  \begin{tabular}{@{}lll@{}}
    \toprule
    \textbf{Measure} & \textbf{What it captures} & \textbf{RQ} \\
    \midrule
    Decision transparency & Visibility and understanding of consequential agent choices & RQ1 \\
    Provenance tracing  & Tracing findings to the choices and evidence that produced them   & RQ1 \\
    Goal connection      & Relating agent findings and choices to the research question & RQ1 \\
    Action confidence & Confidence in defending one's understanding and deciding what to build on & RQ1 \\
    Artifact sensemaking & Comparing outputs and identifying patterns and relationships across them & RQ2 \\
    Orientation support    & Moving between overview/detail and maintaining one's place & RQ3 \\
    NASA-TLX & Subjective workload during review & RQ3 \\
    \bottomrule
  \end{tabular}
  }
\end{table*}

The measures map onto the research questions rather than one-to-one onto the formative challenges. Decision transparency, provenance tracing, goal connection, and action confidence contribute to RQ1; artifact sensemaking addresses RQ2; and orientation support and NASA-TLX address RQ3.

\begin{table*}[!tp]
  \centering
  \small
\caption{Post-task survey items. 7-point Likert scale questions (Strongly Disagree - Strongly Agree). Administered identically after both task blocks.}
  \label{tab:post-task-items}
  \resizebox{0.87\linewidth}{!}{
  \begin{tabular}{@{}lp{13.2cm}@{}}
    \toprule
    \textbf{ID} & \textbf{Item} \\
    \midrule
    \multicolumn{2}{@{}l}{\textit{\textsc{dec} --- Decision transparency}} \\
    \textsc{dec}1 & I could tell which choices the agent made during the run. \\
    \textsc{dec}2 & I could tell why the agent made the choices it made. \\
    \textsc{dec}3 & I could tell which directions the agent considered but did not pursue. \\
    \addlinespace
    \multicolumn{2}{@{}l}{\textit{\textsc{art} --- Artifact sensemaking}} \\
    \textsc{art}1 & I could see what the run produced as a whole. \\
    \textsc{art}2 & I could compare the run's outputs against one another. \\
    \textsc{art}3 & I could identify patterns across the run's outputs. \\
    \textsc{art}4 & The way the outputs were organized helped me identify relationships across the run. \\
    \addlinespace
    \multicolumn{2}{@{}l}{\textit{\textsc{goal} --- Goal connection}} \\
    \textsc{goal}1 & I could tell whether what the agent found addressed the research question. \\
    \textsc{goal}2 & I could tell whether the agent's choices served the research question. \\
    \addlinespace
    \multicolumn{2}{@{}l}{\textit{\textsc{prov} --- Provenance}} \\
    \textsc{prov}1 & I could trace an important finding back to the choices and evidence that produced it. \\
    \addlinespace
    \multicolumn{2}{@{}l}{\textit{\textsc{nav} --- Orientation support}} \\
    \textsc{nav}1 & I could easily move between an overall view of the run and specific details. \\
    \textsc{nav}2 & I could keep track of the run without having to hold too much information in my head. \\
    \addlinespace
    \multicolumn{2}{@{}l}{\textit{\textsc{trust} --- Action confidence}} \\
    \textsc{trust}1 & I could defend my understanding of this run if a colleague asked how I know. \\
    \textsc{trust}2 & I would be comfortable building my next step on what this run produced. \\
    \addlinespace
    \multicolumn{2}{@{}l}{\textit{Single-item global and open response}} \\
    \textsc{glob}1 & Overall, how well did this interface support reviewing the agent's run? (1--7) \\
    \bottomrule
  \end{tabular}
  }
\end{table*}

\begin{table*}[!tp]
\centering
\caption{Post-study comparative survey questions. Participants compared the Chat interface (C) and Chat+Visual interface (CV) based on their experiences reviewing completed agent runs.}
\label{tab:final-survey-comparison}
\small
\resizebox{\linewidth}{!}{
\begin{tabularx}{\textwidth}{
    >{\raggedright\arraybackslash}p{0.04\textwidth}
    >{\raggedright\arraybackslash}p{0.36\textwidth}
    >{\raggedright\arraybackslash}p{0.55\textwidth}}
\toprule
\textbf{\#} & \textbf{Question} & \textbf{Response options} \\
\midrule

1 &
Telling what the agent decided and why? &
Strongly favors C;
Moderately favors C;
Slightly favors C;
No difference;
Slightly favors CV;
Moderately favors CV;
Strongly favors CV
\\

\addlinespace
2 &
Making sense of what the run produced, including relationships across its outputs? &
Strongly favors C;
Moderately favors C;
Slightly favors C;
No difference;
Slightly favors CV;
Moderately favors CV;
Strongly favors CV
\\

\addlinespace
3 &
Telling whether the run served your research question? &
Strongly favors C;
Moderately favors C;
Slightly favors C;
No difference;
Slightly favors CV;
Moderately favors CV;
Strongly favors CV
\\

\addlinespace
4 &
Staying oriented while moving between the overall run and specific details? &
Strongly favors C;
Moderately favors C;
Slightly favors C;
No difference;
Slightly favors CV;
Moderately favors CV;
Strongly favors CV
\\

\addlinespace
5 &
Deciding what parts of the run you would be willing to build on or act on? &
Strongly favors C;
Moderately favors C;
Slightly favors C;
No difference;
Slightly favors CV;
Moderately favors CV;
Strongly favors CV
\\

\addlinespace
6 &
Reviewing the completed run overall? &
Strongly favors C;
Moderately favors C;
Slightly favors C;
No difference;
Slightly favors CV;
Moderately favors CV;
Strongly favors CV
\\

\bottomrule
\end{tabularx}
}
\end{table*}

\begin{table*}[!tp]
\centering
\caption{Comparative survey questions. Responses were given on a seven-point comparative scale: Strongly favors C, Moderately favors C, Slightly favors C,No difference, Slightly favors CV, Moderately favors CV, and Strongly favors CV.}
\label{tab:final-survey-comparison-compact}
\small
\begin{tabularx}{\columnwidth}{
    >{\raggedright\arraybackslash}p{0.06\columnwidth}
    >{\raggedright\arraybackslash}X}
\toprule
\textbf{\#} & \textbf{Which interface better supported you in...} \\
\midrule
1 & Telling what the agent decided and why? \\
\addlinespace
2 & Making sense of what the run produced, including relationships across its outputs? \\
\addlinespace
3 & Telling whether the run served your research question? \\
\addlinespace
4 & Staying oriented while moving between the overall run and specific details? \\
\addlinespace
5 & Deciding what parts of the run you would be willing to build on or act on? \\
\addlinespace
6 & Reviewing the completed run overall? \\
\bottomrule
\end{tabularx}
\end{table*}